\pdfoutput=1
\documentclass[PAPER, atlasdraft=false, coverpage=false, UKenglish, cernpreprint, texmf, orcidlogo]{atlasdoc}
\usepackage{atlaspackage}
\usepackage{atlasbiblatex}

\usepackage{atlasphysics}
\graphicspath{{logos/}{figures/}}

\usepackage{ANA-HIGG-2022-07-PAPER-defs}
\usepackage{multirow}

\AtlasTitle{Differential cross-section measurements of Higgs boson production in the $H\to\tau^+\tau^-$ decay channel in $pp$ collisions at $\sqrt{s}=13$~TeV with the ATLAS detector}

\AtlasAbstract{%

Differential measurements of Higgs boson production in the $\tau$-lepton-pair decay channel are presented in the gluon fusion, vector-boson fusion (VBF), $VH$ and $t\bar{t}H$ associated production modes, with particular focus on the VBF production mode.
The data used to perform  the measurements correspond to 140\,fb$^{-1}$ of proton--proton collisions collected by the ATLAS experiment at the LHC.
Two methods are used to perform the measurements: the \textit{Simplified Template Cross-Section} (STXS) approach and an \textit{Unfolded Fiducial Differential} measurement considering only the VBF phase space.
For the STXS measurement,  events are categorized by their production mode and kinematic properties such as the Higgs boson's transverse momentum ($p^{\text{H}}_\text{T}$), the number of jets produced in association with the Higgs boson,  or the invariant mass of the two leading jets ($m_{jj}$).  For the VBF production mode, the ratio of the measured cross-section to the Standard Model prediction for $m_{jj}>1.5$~TeV and $p^{\text{H}}_\text{T}>200$~GeV ($p^{\text{H}}_\text{T}<200$~GeV) is ${1.29}^{+0.39}_{-0.34}$ (${0.12}^{+0.34}_{-0.33}$). This is the first VBF measurement for the higher-$p^{\text{H}}_\text{T}$~criteria, and the most precise for the lower-$p^{\text{H}}_\text{T}$~criteria.
The \textit{fiducial} cross-section measurements,  which only  consider the kinematic properties of the event,  are performed as functions of variables  characterizing the VBF topology, such as the signed $\Delta\phi_{jj}$ between the two leading jets.  The measurements have a precision of 30\%--50\% and agree well with the Standard Model predictions.
These results are interpreted in the SMEFT framework, and place the strongest constraints to date on the CP-odd Wilson coefficient $c_{H\tilde{W}}$.
}

\AtlasRefCode{HIGG-2022-07}

\PreprintIdNumber{CERN-EP-2024-198}

\AtlasJournalRef{JHEP 03 (2025) 010}

\AtlasDOI{10.1007/JHEP03(2025)010}

\hypersetup{pdftitle={ATLAS document},pdfauthor={The ATLAS Collaboration}}

\begin{document}

\maketitle

\tableofcontents

\section{Introduction}
%

Since the discovery of the Higgs boson by the ATLAS~\cite{HIGG-2012-27} and CMS~\cite{CMS-HIG-12-028} Collaborations in 2012, studying the properties of this boson has been a priority of the LHC experiments.
The main Higgs boson production mode at the LHC is gluon--gluon fusion (ggF), followed by vector boson fusion (VBF) production, associated production with a vector boson ($VH$, $V= W$ or $Z$) and associated production with a top-antitop quark pair (\ttH).
In the Standard Model (SM), the decay into a $\tau^+\tau^-$ pair\footnote{For simplicity, a $\tau^+\tau^-$~pair is denoted by $\tau\tau$~throughout this paper.} has the largest branching ratio of all leptonic Higgs boson decays.

Previously, the ATLAS Collaboration used the \RunTwo dataset to study the $H\to\tau\tau$ final state~\cite{HIGG-2019-09} and three measurements were performed: the total Higgs boson production cross-section, the cross-sections for the four dominant production modes, and a measurement of the cross-section in nine phase-space regions (\emph{bins}) in the Simplified Template Cross-Section (STXS) framework~\cite{deFlorian:2016spz}. For the STXS measurement, cross-sections for Higgs boson production were measured as a function of the transverse momentum (\pT) of the Higgs boson (\pTH), the number of jets produced in association with the Higgs boson, and the invariant mass of the two leading jets ($\mjj$) where applicable. Higgs boson production through VBF was confirmed in this decay channel with an observed (expected) significance of 5.3$\sigma$ (6.2$\sigma$) over the SM background, and a cross-section of $0.90{}^{+0.20}_{-0.17}$ times the SM prediction. Of all the Higgs boson decay channels explored by ATLAS~\cite{HIGG-2019-09,HIGG-2018-28,HIGG-2020-16,HIGG-2021-20}, $H\to\tau\tau$ provided the most precise single measurement of VBF production and motivates further studies of the kinematics of Higgs boson production using this channel.  The results in this paper constitute the final \RunTwo measurements in the $H\to\tau\tau$ final state and supersede those in Ref.~\cite{HIGG-2019-09}.
A similar measurement was also performed by the CMS Collaboration using the \RunTwo dataset \cite{CMS-HIG-19-010}, where the cross-section for the Higgs boson production through VBF was measured as $0.81 \pm 0.17$ times the SM prediction.

Two approaches for investigating the properties of VBF events in the $H\rightarrow\tau\tau$ decay channel  are reported in this paper.
The first approach is to perform a cross-section measurement within the STXS framework, expanding on the measurement of Ref.~\cite{HIGG-2019-09} in which only a single bin in the VBF phase space was measured.
In this context, the extracted cross-sections are based on the production mode and the kinematics of the events; in the particular case of VBF production, the events are separated into bins based on \pTH and $\mjj$.
Similar measurements were also pursued for other Higgs boson decay channels such as $H\rightarrow ZZ^*$~\cite{HIGG-2018-28,CMS-HIG-19-001}, $H\rightarrow \gamma\gamma$~\cite{HIGG-2020-16,CMS-HIG-19-015} and $H\rightarrow WW^*$~\cite{HIGG-2021-20,CMS-HIG-20-013}.
The second approach is to measure \textit{fiducial} differential production cross-sections for \Htautau events. The fiducial measurement approach does not distinguish between the different Higgs boson production modes, but a phase-space region enriched in VBF events is defined to ensure optimal measurement sensitivity.
This results in a less model-dependent approach than the STXS framework, although still relying on simulated SM samples to derive response matrices.
The cross-section is measured as a function of several variables, including \pTH, the azimuthal angle $\dPhiJJsign$ between the jets, and the leading jet's transverse momentum, \pTj. They are chosen to test the kinematics of the SM Higgs boson and the VBF production mode, as well as to provide sensitivity to contributions from physics beyond the SM.
Differential measurements in the VBF phase space were performed previously in the $H\rightarrow\gamma\gamma$~\cite{HIGG-2019-13} and $H\rightarrow WW^*$~\cite{HIGG-2020-25} channels, while this is the first differential measurement performed in the \Htautau channel.

Additionally, the results reported in this paper extend the study of \ttH production in the STXS framework.
In Ref.~\cite{HIGG-2019-09}, an inclusive measurement of \ttH production was performed in the fully hadronic final state, whereas this paper considers multiple bins in \pTH because higher momentum bins are expected to be more sensitive to physics beyond the SM. %

This paper is organized as follows. The ATLAS detector is described in Section~\ref{sec:detector}, followed by the data and simulated samples in Section~\ref{sec:samples}, and the object reconstruction and main event selection in Section~\ref{sec:analysis}.
Section~\ref{sec:stxs} discusses the event categorization, and then Section~\ref{sec:background} delves into more details of the background estimation methods.
The systematic uncertainties are discussed in Section~\ref{sec:systematic}, followed by the STXS cross-section measurement results in Section~\ref{sec:stxs_results} and the differential cross-section measurement results in Section~\ref{sec:differential_results}.
Conclusions are reported in Section~\ref{sec:conclusion}.

%


%
\section{ATLAS detector}
\label{sec:detector}

%
\newcommand{\AtlasCoordFootnote}{%
ATLAS uses a right-handed coordinate system with its origin at the nominal interaction point (IP)
in the centre of the detector and the \(z\)-axis along the beam pipe.
The \(x\)-axis points from the IP to the centre of the LHC ring,
and the \(y\)-axis points upwards.
Polar coordinates \((r,\phi)\) are used in the transverse plane,
\(\phi\) being the azimuthal angle around the \(z\)-axis.
The pseudorapidity is defined in terms of the polar angle \(\theta\) as \(\eta = -\ln \tan(\theta/2)\) and is equal to the rapidity
$ y = \frac{1}{2} \ln \left( \frac{E + p_z c}{E - p_z c} \right) $ in the relativistic limit.
Angular distance is measured in units of \(\Delta R \equiv \sqrt{(\Delta y)^{2} + (\Delta\phi)^{2}}\).}

The ATLAS detector~\cite{PERF-2007-01} at the LHC covers nearly the entire solid angle around the collision point.\footnote{\AtlasCoordFootnote}
It consists of an inner tracking detector surrounded by a thin superconducting solenoid, electromagnetic and hadronic calorimeters,
and a muon spectrometer incorporating three large superconducting air-core toroidal magnets.

The inner-detector system (ID) is immersed in a \qty{2}{\tesla} axial magnetic field
and provides charged-particle tracking in the range \(|\eta| < 2.5\).
The high-granularity silicon pixel detector covers the vertex region and typically provides four measurements per track,
the first hit generally being in the insertable B-layer (IBL) installed before Run~2~\cite{ATLAS-TDR-19,PIX-2018-001}.
It is followed by the SemiConductor Tracker (SCT), which usually provides eight measurements per track.
These silicon detectors are complemented by the transition radiation tracker (TRT),
which enables radially extended track reconstruction up to \(|\eta| = 2.0\).
The TRT also provides electron identification information
based on the fraction of hits (typically 30 in total) above a higher energy-deposit threshold corresponding to transition radiation.

The calorimeter system covers the pseudorapidity range \(|\eta| < 4.9\).
Within the region \(|\eta|< 3.2\), electromagnetic calorimetry is provided by barrel and
endcap high-granularity lead/liquid-argon (LAr) calorimeters,
with an additional thin LAr presampler covering \(|\eta| < 1.8\)
to correct for energy loss in material upstream of the calorimeters.
Hadronic calorimetry is provided by the steel/scintillator-tile calorimeter,
segmented into three barrel structures within \(|\eta| < 1.7\), and two copper/LAr hadronic endcap calorimeters.
The solid angle coverage is completed with forward copper/LAr and tungsten/LAr calorimeter modules
optimised for electromagnetic and hadronic energy measurements respectively.

The muon spectrometer (MS) comprises separate trigger and
high-precision tracking chambers measuring the deflection of muons in a magnetic field generated by the superconducting air-core toroidal magnets.
The field integral of the toroids ranges between \num{2.0} and \qty{6.0}{\tesla\metre}
across most of the detector.
Three layers of precision chambers, each consisting of layers of monitored drift tubes, cover the region \(|\eta| < 2.7\),
complemented by cathode-strip chambers in the forward region, where the background is highest.
The muon trigger system covers the range \(|\eta| < 2.4\) with resistive-plate chambers in the barrel, and thin-gap chambers in the endcap regions.

The luminosity is measured mainly by the LUCID--2~\cite{LUCID2} detector that records Cherenkov light produced in the quartz windows of photomultipliers located close to the beampipe.

Events are selected by the first-level trigger system implemented in custom hardware,
followed by selections made by algorithms implemented in software in the high-level trigger~\cite{TRIG-2016-01}.
The first-level trigger accepts events from the \qty{40}{\MHz} bunch crossings at a rate below \qty{100}{\kHz},
which the high-level trigger further reduces in order to record complete events to disk at about \qty{1}{\kHz}.

A software suite~\cite{SOFT-2022-02} is used in data simulation, in the reconstruction
and analysis of real and simulated data, in detector operations, and in the trigger and data acquisition
systems of the experiment.


%
\section{Data and simulated event samples}
\label{sec:samples}

The data used to perform this analysis correspond to proton--proton collisions at a centre-of-mass energy $\sqrt{s}=13$~TeV collected by the ATLAS detector during \RunTwo of the LHC (2015--2018). %
Events are selected for analysis only if they pass a good-quality requirement and all relevant detector components were in good operating condition~\cite{DAPR-2018-01}. This yields a total integrated luminosity of 140.1\,fb$^{-1}$~\cite{DAPR-2021-01}.
The uncertainty in the integrated luminosity is 0.83\%~\cite{DAPR-2021-01}, obtained using the LUCID-2 detector \cite{LUCID2} for the primary luminosity measurements, complemented by measurements using the inner detector and calorimeters.

Monte Carlo (MC) simulated event samples are used to model the \Htautau signal and most of the backgrounds from SM processes.
A summary of the generators used for the simulation is shown in \cref{tab:generators}.
The same MC samples were also used in Ref.~\cite{HIGG-2019-09}, and only a brief description of the main samples is reported in this section.
All MC samples were reconstructed in the same manner as the data.

\begin{table}[htb]
\caption{Summary of the MC generators used for the main signal and background samples. Information about the matrix element (ME) generator, parton shower (PS), parton distribution function (PDF) set used in the ME and the PS, and set of tuned parameter (Tune) is provided. The column labelled `Normalization' specifies the order of the cross-section calculation used for the normalization of the simulated samples.}
\label{tab:generators}
\resizebox{\textwidth}{!}{
\begin{tabular}{l c c c c l l}
\toprule
Process               & \multicolumn{2}{c}{Generator}                                   & \multicolumn{2}{c}{PDF set}                                      & Tune  & Normalization        \\
& ME                                & PS                          & ME                                        & PS                   &       &              \\
\midrule
Higgs boson     &                                   &                             &                                           &                      &                      \\
\midrule
ggF                   & \POWHEGBOX[v2]                           & \PYTHIA[8]                    & \PDFforLHC[15nnlo]                & \CTEQ[6L1]              & AZNLO & N$^{3}$LO QCD + NLO EW    \\
VBF                   & \POWHEGBOX[v2]                           & \PYTHIA[8]                    & \PDFforLHC[15nlo]                                & \CTEQ[6L1]              & AZNLO & NNLO QCD + NLO EW        \\
$VH$                  & \POWHEGBOX[v2]                           & \PYTHIA[8]                    & \PDFforLHC[15nlo]                               & \CTEQ[6L1]              & AZNLO & NNLO QCD + NLO EW          \\
\ttH                  & \POWHEGBOX[v2]                        & \PYTHIA[8]                    & \NNPDF[3.0nnlo] & \NNPDF[2.3lo]              & A14                  & NLO QCD + NLO EW                  \\
\midrule
Background            &                                   &                             &                                           &                      &                      \\
\midrule
\Vjets (QCD/EW)      & \multicolumn{2}{c}{\SHERPA[2.2.1]} & \multicolumn{2}{c}{\NNPDF[3.0nnlo]} & \SHERPA                                   & NNLO for QCD, LO for EW                                       \\
\ttbar                & \POWHEGBOX[v2]                           & \PYTHIA[8]                    & \NNPDF[3.0nnlo] & \NNPDF[2.3lo]               & A14                  & NNLO + NNLL QCD        \\
\bottomrule
\end{tabular}
}
\end{table}

The dominant Higgs boson production mode at the LHC is ggF, with a cross-section of \SI{48.6}{\pico\barn},\footnote{The cross-sections are reported for a centre-of-mass energy of $\sqrt{s}=13$~TeV.}  followed by VBF production, with a cross-section of \SI{3.78}{\pico\barn}, then $VH$ production,  with a cross-section of \SI{2.25}{\pico\barn}, and \ttH production, with a cross-section of \SI{0.51}{\pico\barn}.
Higgs boson production via ggF was simulated using the \POWHEGBOX[v2]~\cite{Frixione:2007nw,Nason:2004rx,Frixione:2007vw,Alioli:2010xd,Hartanto:2015uka} generator at next-to-next-to-leading-order (NNLO) accuracy in QCD. The MC simulated samples are normalized to the next-to-next-to-next-to-leading-order (N${}^3$LO) cross-section in QCD plus electroweak (EW) corrections at next-to-leading order (NLO)~\cite{deFlorian:2016spz,Anastasiou:2016cez,Anastasiou:2015ema,Dulat:2018rbf,Harlander:2009mq,Harlander:2009bw,Harlander:2009my,Pak:2009dg,Actis:2008ug,Actis:2008ts,Bonetti:2018ukf,Bonetti:2018ukf}. For this process the \PDFforLHC[15nnlo] set of parton distribution functions (PDF)~\cite{Butterworth:2015oua} was used, while the parton shower (PS) and non-perturbative effects were modelled with \PYTHIA[8.230]~\cite{Sjostrand:2014zea} with parameter values set to those of the \AZNLO~\cite{STDM-2012-23} set of tuned parameters (tune).

Higgs boson production via VBF was simulated using \POWHEGBOX[v2] at NLO accuracy in QCD.
It was tuned to match calculations with effects due to finite heavy-quark masses and soft-gluon resummation up to next-to-next-to-leading logarithms (NNLL).
The MC simulated samples are normalized to an approximate-NNLO QCD cross-section with NLO electroweak corrections~\cite{Ciccolini:2007jr,Ciccolini:2007ec,Bolzoni:2010xr}. The \PDFforLHC[15nlo] PDF set was used, while \PYTHIA[8.230] with the \AZNLO tune was used for PS simulation.

Higgs boson production via $VH$ was simulated with \POWHEGBOX[v2] at NLO accuracy in QCD for $VH$ plus one-jet production.
The loop-induced $gg\to ZH$ process was generated separately at leading order (LO) in QCD.
The MC simulated samples are normalized to cross-sections calculated at NNLO in QCD with NLO electroweak corrections for $p p \to VH$ and at NLO and next-to-leading-logarithm accuracy in QCD for $gg \to ZH$~\cite{Ciccolini:2003jy,Brein:2003wg,Brein:2011vx,Altenkamp:2012sx,Denner:2014cla,Brein:2012ne,Harlander:2014wda}.
Similarly to the VBF samples, the \PDFforLHC[15nlo] PDF set was used, while \PYTHIA[8.230] with the \AZNLO tune was used for PS simulation.

The production of \ttH events was simulated at NLO accuracy in QCD using \POWHEGBOX[v2], with the decays of bottom and charm hadrons performed by \EVTGEN[1.6.0]~\cite{Lange:2001uf}.
The cross-section used to normalize the \ttH process is calculated at NLO in QCD and electroweak couplings~\cite{deFlorian:2016spz,Beenakker:2002nc,Dawson:2003zu,Yu:2014cka,Frixione:2014qaa}. For this sample, the \NNPDF[3.0nlo] PDF set~\cite{Ball:2014uwa} was used, with the PS modelled by \PYTHIA[8.230] with the A14 tune~\cite{ATL-PHYS-PUB-2014-021}.

The normalization of all Higgs boson samples accounts for the decay branching ratio calculated with HDECAY~\cite{Djouadi:1997yw,Spira:1997dg,Djouadi:2006bz} and \PROPHECY~\cite{Bredenstein:2006ha,Bredenstein:2006rh,Bredenstein:2006nkV2}. The decays and spin correlations for $\tau$-leptons are handled by \PYTHIA. %
The SM prediction of the \Htautau branching ratio ($0.0625 \pm 0.0014$) and a Higgs boson mass of \SI{125.09}{\GeV} are assumed in the calculation of the expected cross-sections throughout this measurement.

The QCD production of the \Vjets background was simulated with the \SHERPA[2.2.1]~\cite{Bothmann:2019yzt} generator using NLO matrix elements for up to two partons, and LO matrix elements for up to four partons, calculated with the \textsc{Comix}~\cite{Gleisberg:2008fv} and \OPENLOOPS~\cite{Buccioni:2019sur,Cascioli:2011va,Denner:2016kdg} libraries.
They were matched with the \SHERPA parton shower~\cite{Schumann:2007mg} using the MEPS@NLO prescription~\cite{Hoeche:2011fd,Hoeche:2012yf,Catani:2001cc,Hoeche:2009rj} using the set of tuned parameters developed by the \SHERPA authors.
The \NNPDF[3.0nnlo] set of PDFs~\cite{Ball:2014uwa} was used and the samples are normalized to a NNLO prediction~\cite{Anastasiou:2003ds}.
Electroweak production of the $lljj$ final state (with $l=e, \mu,\tau$, and $j$ denoting a jet) was simulated with \SHERPA[2.2.1] using LO matrix elements with up to two additional parton emissions. The LO cross-section is multiplied by a factor 1.7 to match the cross-section value measured in Ref.~\cite{STDM-2016-09}; this scaling was also applied in Ref.~\cite{HIGG-2018-14}.
For the matching with the parton shower, the same prescription as for the QCD \Vjets processes was used.
The samples were produced using the VBF approximation, which avoids overlap with semileptonic diboson topologies by requiring a $t$-channel colour-singlet exchange.

The production of \ttbar\ events was modelled at NLO in QCD by \POWHEGBOX[v2] with the \NNPDF[3.0nlo] PDF set and with the \hdamp parameter\footnote{The \hdamp\ parameter is a resummation damping factor and one of the parameters that controls the matching of \POWHEG matrix elements to the parton shower and thus effectively regulates the high-\pt radiation against which the \ttbar\ system recoils.} set to 1.5\,\mtop~\cite{ATL-PHYS-PUB-2016-020}.
The events were interfaced to \PYTHIA[8.230] to model the parton shower, hadronization, and underlying event, with parameters set according to the A14 tune and using the \NNPDF[2.3lo] set of PDFs.
The decays of bottom and charm hadrons were performed by \EVTGEN, as they were for the \ttH sample.
The \ttbar sample is normalized to the cross-section prediction at NNLO in QCD, including the resummation of NNLL soft-gluon terms, calculated using \TOPpp[2.0]~\cite{Beneke:2011mq,Cacciari:2011hy,Baernreuther:2012ws,Czakon:2012zr,Czakon:2012pz,Czakon:2013goa,Czakon:2011xx}.

To simulate the effects of additional $pp$ collisions in the same, next or previous bunch crossing (pile-up), additional interactions were generated using the soft QCD processes provided by \PYTHIA[8.186]~\cite{Sjostrand:2007gs} with the A3 tune~\cite{ATL-PHYS-PUB-2016-017} and the MSTW2008LO PDF set~\cite{Martin:2009iq}, and overlaid onto each simulated hard-scatter event.
The MC samples were reweighted so that the pile-up distribution matches the one observed in the data, and were processed through an ATLAS detector simulation~\cite{SOFT-2010-01} based on \GEANT~\cite{Agostinelli:2002hh}.


%
\section{Object reconstruction and event selection}
\label{sec:analysis}

The selection of \Htautau events requires the reconstruction of electrons, muons, the visible products of hadronically decaying $\tau$-leptons (\tauhadvis), jets (including their $b$-tagging properties) and missing transverse momentum.
The numbers of reconstructed light leptons (electrons and muons, globally referred to as `leptons' or $\ell$) and \tauhadvis in each event are used for its assignment to one of the three channels used in the analysis.
Requirements on the number of additional jets in the event are then used in the signal region categorization and for background suppression.
Both the object selection criteria and the basic event selection are taken from Ref.~\cite{HIGG-2019-09}.

Three channels from the \Htautau decay are considered: two hadronically decaying $\tau$-leptons (\tauhad, where the $\tau$-lepton decays into hadrons plus a neutrino), denoted by \thadhad; one leptonically decaying $\tau$-lepton (\taulep) and one \tauhad, denoted by \tlhad; and two \taulep with different flavours, denoted by \temu.
Events were collected with single-lepton,  dilepton or di-hadronic $\tau$-lepton triggers~\cite{TRIG-2018-05,TRIG-2018-01,TRIG-2019-03,TRIG-2019-02}.
The requirements imposed on the \pt of the reconstructed electrons, muons or \tauhadvis in the analysis are tighter than those applied in the trigger-object reconstruction and trigger decision,
in order to be on the trigger efficiency plateau.

Electrons are reconstructed from topological clusters of energy deposits in the electromagnetic calorimeter which are matched to a track reconstructed in the ID~\cite{ATLAS:2023dxj}.
They are required to satisfy the \objectsc{Loose} identification criteria described in Ref.~\cite{ATLAS:2023dxj}, to have $\pT>\SI{15}{\GeV}$, and to be in the fiducial volume of the ID and the high-granularity electromagnetic calorimeters, $|\eta_\text{cluster}|<2.47$, with the exclusion of the LAr barrel--endcap transition region $1.37 < |\eta_\text{cluster}| < 1.52$.
In the \temu and \tehad channels, the selected electron is also required to satisfy the \objectsc{Medium} identification and the \objectsc{Loose} isolation criterion~\cite{ATLAS:2023dxj} in the signal regions and in most of the control regions. The \objectsc{Medium} identification has an efficiency of \SI{80}{\percent} to \SI{90}{\percent}, while \objectsc{Loose} isolation has an efficiency of \SI{90}{\percent} for candidates with $\pT=$ \SI{15}{\GeV}, and it increases to more than \SI{98}{\percent} for candidates with $\pT=$ \SI{30}{\GeV}.
The \pT requirement is tightened if the event is accepted by the single-electron trigger or the electron--muon trigger, as shown in \cref{tab:selection:ptcut}.

Muons are reconstructed from signals in the MS matched with tracks in the ID.
They are required to satisfy the \objectsc{Loose} identification criteria~\cite{MUON-2018-03} and to have $\pT > \SI{10}{\GeV}$ and $|\eta|<2.5$.
In the \temu and \tmhad channels, the selected muon in the signal regions is also required to satisfy the \objectsc{Medium} identification and the \objectsc{Tight} isolation criterion~\cite{MUON-2018-03} based on track information. The \objectsc{Medium} identification has an efficiency above \SI{97}{\percent}, while the \objectsc{Tight} isolation has an efficiency increasing from \SI{85}{\percent} to \SI{99}{\percent} for muons with $\pT$ increasing from \SI{10}{\GeV} to \SI{50}{\GeV}.
The \pT requirement is tightened if the event is accepted by the single-muon trigger or the electron--muon trigger, as shown in \cref{tab:selection:ptcut}.

\begin{table}[t]%
\caption{
Transverse momentum thresholds applied to the selected electrons, muons and $\tau_{\text{had-vis}}$  depending on the trigger signature and the data-taking period.
The \pT thresholds of the lowest-\pT unprescaled triggers during the \RunTwo ATLAS data-taking are reported in Refs.~\cite{ATL-DAQ-PUB-2016-001,ATL-DAQ-PUB-2017-001,ATL-DAQ-PUB-2018-002,ATL-DAQ-PUB-2019-001}.
}
\label{tab:selection:ptcut}
\begin{center}
\scalebox{0.9}{
\begin{tabular}{@{}lcc@{}}
\toprule
Trigger signature              & Data-taking period                & \pT threshold [GeV] used in event selection    \\
\midrule
\multirow{2}{*}{Single electron} & 2015                       & $\pT(e)>25$                     \\
& 2016--2018                  & $\pT(e)>27$                      \\
\midrule
\multirow{2}{*}{Single muon}     & 2015                       & $\pT(\mu)>21$      \\
& 2016--2018                  & ~~~$\pT(\mu)>27.3$                \\
\midrule
One electron, one muon                    & 2015--2018                  & $\pT(e)>18$, $\pT(\mu)>14.7$ \\
\midrule
\multirow{2}{*}{Two \tauhadvis}           & \multirow{2}{*}{2015--2018} & ~~~~~~~$\pT(\textrm{leading}~\tauhadvis)>40$     \\
&                            & $\pT(\textrm{sub-leading}~\tauhadvis)>30$  \\
\bottomrule
\end{tabular}
}
\end{center}
\end{table}

Jets are reconstructed by applying the \antikt algorithm~\cite{Cacciari:2008gp,Fastjet} with a radius parameter of $R=0.4$ to
topological calorimeter-energy clusters and charged-particle tracks processed with a particle-flow algorithm~\cite{PERF-2015-09}.
The energy of the jets is corrected for the non-compensating calorimeter response, noise threshold effects, energy loss from inactive material, and pile-up contamination~\cite{JETM-2018-05}.
Only jets with $\pT>\SI{20}{\GeV}$ and $|\eta|<4.5$ are considered.
A jet vertex tagger (JVT)~\cite{PERF-2014-03} is used to remove jets with $\pT<\SI{60}{\GeV}$ and $\abseta<2.5$ that are identified as not being associated with the event's primary vertex.
Similarly, pile-up jets in the forward region are suppressed with a \enquote{forward JVT}~\cite{PERF-2016-06} algorithm, which exploits jet shapes and topological jet correlations in pile-up interactions and is applied to all jets with $\pT<\SI{60}{\GeV}$ and $|\eta|>2.5$.
Jets with $\pT>\SI{20}{\GeV}$ and $|\eta|<2.5$ containing $b$-hadrons are identified using the DL1r $b$-tagging algorithm~\cite{FTAG-2019-07}. Two different working points (WP) for $b$-tagging are used depending on the channel: one with \SI{85}{\percent} efficiency is used in the \temu and \tlhad channels, while another with \SI{70}{\percent} efficiency is used in the \thadhad channel.
The rejection factors for $b$-tagging of jets initiated by $c$-quarks and light partons are 9.4 (2.6) and 390 (29) respectively for the \SI{70}{\percent} (\SI{85}{\percent}) efficiency working point.

Hadronic $\tau$-lepton decays produce a neutrino and visible decay products, mostly one or three charged pions and mostly up to two neutral pions.
The reconstruction of the \tauhadvis is seeded by jets reconstructed using the \antikt algorithm with a radius parameter of $R=0.4$. %
Reconstructed tracks are then matched to \tauhadvis candidates.
A multivariate discriminant is used to identify the tracks which are likely to have been produced by the charged \tauhad decay products, while rejecting tracks originating from other interactions, nearby jets, photon conversions or misreconstructed tracks~\cite{ATL-PHYS-PUB-2015-045}.
The \tauhadvis objects are required to have one or three associated tracks selected by using this discriminant.
Their charge ($q$) is defined as the sum of the measured charges of these associated tracks and is required to be $|q|=1$.
The \tauhadvis objects must also satisfy the requirements $\pT>\SI{20}{\GeV}$ and $|\eta|< 2.47$, excluding the region $1.37 < |\eta| < 1.52$.
The \pT requirement is tightened if the event is accepted by a di-hadronic $\tau$-trigger, as shown in \cref{tab:selection:ptcut}.
In order to separate the \tauhadvis candidates produced by hadronic $\tau$ decays from those produced by jets initiated by quarks or gluons, a recurrent neural-network  identification algorithm~\cite{ATL-PHYS-PUB-2019-033} is used.  This algorithm uses information from reconstructed charged-particle tracks and calorimeter energy clusters associated with \tauhadvis candidates, together with the values of high-level discriminating variables. In the signal regions of the \tlhad and \thadhad channels, the \tauhadvis objects are required to fulfil the \objectsc{Medium} identification criterion, which has an efficiency of 75\% (60\%) for candidates with one~(three) associated track(s).
A separate discriminant (eBDT) is constructed to reject backgrounds originating from electrons misidentified as \tauhadvis, arising mostly from \Zee events in this analysis.
In the \tehad channel, for events where only one charged track is matched to the \tauhadvis object, the \tauhadvis is required to pass the \objectsc{Medium} working point of the eBDT algorithm, which has an 85\% efficiency for candidates satisfying the identification requirement.

Since the objects used in the analysis are reconstructed from the same set of tracks and calorimetric energy clusters, some constituents can be associated with multiple objects.
These overlaps between different objects are resolved with the procedure described in Ref.~\cite{HIGG-2019-09}.

Simulated events are corrected for differences between data and MC simulation seen in jet vertex tagger and $b$-tagging efficiencies as well as $b$-tagging mis-tag rates~\cite{PERF-2014-03,FTAG-2018-01,FTAG-2020-08,FTAG-2019-02}. Corrections are also applied to account for minor differences between data and MC simulation seen in the lepton and $\tau$-lepton trigger, reconstruction, identification and isolation efficiencies~\cite{ATLAS:2023dxj,MUON-2018-03,ATLAS-CONF-2017-029}.

The missing transverse momentum vector, $\vecmet$, is reconstructed as the negative vector sum of the transverse momenta of leptons, \tauhadvis and jets, and a soft term, which is calculated as the vectorial sum of the \pT of tracks matched to the primary vertex but not associated with a reconstructed lepton, \tauhadvis or jet~\autocite{JETM-2020-03}.
The magnitude of $\vecmet$ is referred to as the missing transverse energy, \met.

After the object reconstruction and the overlap removal, the events are selected if they contain a $\tau\tau$ pair in one of the channels being studied: \temu, \tlhad or \thadhad.
For each channel, additional selections based on topological criteria such as angular distances, the transverse mass \mT of the lepton and \met defined as $\mT^2=2 \pT^{\ell}\met(1 - \cos(\Delta\phi(\ell,\met)))$, and the fraction of each $\tau$-lepton's momentum carried by its visible decay products in the collinear approximation~\cite{collinear_approximation} ($x_{1}$ and $x_{2}$) are applied to enhance the sensitivity to the Higgs boson signal and reject background events. The collinear approximation assumes that the neutrinos from the $\tau$-lepton decays propagate in the same direction as the visible decay products. Additionally, a requirement of at least one jet in the event is added to the selection for each channel.
These selection criteria are summarized in \cref{tab:selection:baseline}.

The invariant mass ($m_{\tau\tau}$) of the two $\tau$-leptons is calculated using a likelihood-based technique, called the Missing Mass Calculator (MMC)~\cite{MMCpaper}, which relies on information about the $\tau$-lepton candidates' momenta, the presence of additional jets, the \vecmet and the type of $\tau$-lepton decay. In cases where the likelihood does not converge, the invariant mass is estimated in the collinear approximation, denoted by $m_{\tau\tau}^{\mathrm{coll}}$.
The mass reconstruction techniques assume that the missing transverse momentum in the event is solely due to neutrinos from the $\tau$-lepton decays. These mass reconstruction techniques are important in separating the signal from the \Ztt background. In addition, $m_{\tau\tau}^{\mathrm{coll}}$ is used to separate the phase space of the \temu channel from the $H\rightarrow WW^*$ analysis.

\begin{table}[tbhp]
\caption{
Summary of the event selection.
In the $\tau_{e}\tau_{\mu}$ channel, events recorded with the electron trigger must satisfy $\pT(e)>27$\,\GeV\ and $\pT(\mu)>10$\,\GeV, events recorded with the muon trigger must satisfy  $\pT(e)>15$\,\GeV\ and $\pT(\mu)>27.3$\,\GeV, and events recorded with the electron--muon trigger must satisfy $\pT(e)>18$\,\GeV\ and $\pT(\mu)>14.7$\,\GeV.
}
\label{tab:selection:baseline}
\begin{center}
\scalebox{0.8}{
\begin{tabular}{r|c|c|c|c}
\toprule
& \multirow{2}{*}{\temu} & \multicolumn{2}{c|}{\tlhad}  & \multirow{2}{*}{\thadhad} \\
&       & $e\tauhad$ & $\mu\tauhad$              &  \\
\midrule
&
& \multicolumn{2}{c|}{}
&  \\
\textbf{Preselection} &
& \multicolumn{2}{c|}{}
&  \\
Object counting       & $N_{e}$ = 1, $N_{\mu} = 1$, $N_{\tauhadvis} = 0$
& \multicolumn{2}{c|}{ $N_{e/\mu} = 1$, $N_{\tauhadvis} = 1$}
& $N_{e/\mu} = 0$, $N_{\tauhadvis} = 2$\\
&
& \multicolumn{2}{c|}{ }
& \\
\pT cut               & $e$/$\mu$ : \pT cut \SIrange{10}{27.3}{GeV}
& \multicolumn{2}{c|}{$e$/$\mu$ : \pT cut \SIrange{21}{27.3}{GeV},}
& \\
&
& \multicolumn{2}{c|}{\tauhadvis: $\pT > \SI{30}{\GeV}$}
& \tauhadvis: $\pT > \SI{40}, \SI{30}{\GeV}$ \\
&
& \multicolumn{2}{c|}{ }
& \\
ID,                   & $e$/$\mu$: Medium
& \multicolumn{2}{c|}{$e$/$\mu$: Medium, \tauhadvis: Medium}
& \tauhadvis: Medium \\
Isolation,            & $e$: Loose; $\mu$: Tight
& \multicolumn{2}{c|}{$e$: Loose; $\mu$: Tight}
& \\
and $e$-veto             &
& 1-prong \tauhadvis: &
&  \\
&
& \multicolumn{1}{r|}{eBDT $e$-veto} &
& \\
&
& \multicolumn{2}{c|}{ }
& \\
&
& \multicolumn{2}{c|}{ }
& \\
Charge product        & Opposite charge
& \multicolumn{2}{c|}{Opposite charge}
& Opposite charge \\
&
& \multicolumn{2}{c|}{ }
& \\
Kinematics            & $m_{\tau\tau}^{\mathrm{coll}} > m_Z - \SI{25}{\GeV}$
& \multicolumn{2}{c|}{$\mT < \SI{70}{\GeV}$}
& \\
& $30 < m_{e\mu} < \SI{100}{\GeV}$
& \multicolumn{2}{c|}{ }
& \\
&
& \multicolumn{2}{c|}{ }
& \\
$b$-veto              & $N_{b\text{-jets}} = 0$
& \multicolumn{2}{c|}{$N_{b\text{-jets}} = 0$}
& $N_{b\text{-jets}} = 0$ \\
& DL1r 85$\%$ WP
& \multicolumn{2}{c|}{DL1r 85$\%$ WP}
& DL1r 70$\%$ WP \\
&
& \multicolumn{2}{c|}{ }
& not applied in tt(0L)\Hthadthad\\
&
& \multicolumn{2}{c|}{ }
& \\
\MET                 & $\MET > \SI{20}{\GeV}$
& \multicolumn{2}{c|}{$\MET > \SI{20}{\GeV}$}
& $\MET > \SI{20}{\GeV}$\\
&
& \multicolumn{2}{c|}{ }
& \\
Leading jet    & $\pT > \SI{40}{\GeV}$
& \multicolumn{2}{c|}{$\pT > \SI{40}{\GeV}$}
& $\pT > \SI{70}{\GeV}$, $|\eta| < 3.2$ \\
&
& \multicolumn{2}{c|}{ }
& \\
&
& \multicolumn{2}{c|}{ }
& \\
Angular          & $\Delta R_{e\mu} < 2.0$, $|\Delta\eta_{e\mu}| < 1.5$
& \multicolumn{2}{c|}{$\Delta R_{\ell\tauhadvis} < 2.5$, $|\Delta\eta_{\ell\tauhadvis}| < 1.5$}
& $0.6 < \Delta R_{\tauhadvis\tauhadvis} < 2.5$ \\
&
& \multicolumn{2}{c|}{ }
& $|\Delta\eta_{\tauhadvis\tauhadvis}| < 1.5$\\
&
& \multicolumn{2}{c|}{ }
& \\
$x_{1}$ and $x_{2}$         & $0.1 < x_{1} < 1.0$, $0.1 < x_{2} < 1.0$
& \multicolumn{2}{c|}{$0.1 < x_{1} < 1.4$, $0.1 < x_{2} < 1.2$}
& $0.1 < x_{1} < 1.4$, $0.1 < x_{2} < 1.4$ \\
&
& \multicolumn{2}{c|}{}
&  \\
\bottomrule
\end{tabular}
}
\end{center}
\end{table}



%
\section{Event categorization }
\label{sec:stxs}

After the baseline event selection reported in \cref{tab:selection:baseline}, an additional event categorization is performed to select Higgs boson events in different STXS production modes and in a signal region targeting the VBF production mode for the differential fiducial measurement.
The applied requirements are summarized in Table~\ref{tab:selection:signalregions}.
Subsequently, within each STXS signal region, a further event splitting is performed to provide sensitivity to the targeted STXS bins. %
The requirements applied for this splitting are unchanged from those in Ref.~\cite{HIGG-2019-09} for the Boost and VH categories, which target ggF and V(had)H production, respectively.
To target additional STXS bins, finer requirements are made for VBF categories as well for those targeting $\ttH$ production with fully hadronic top-quark and $\tau$-lepton decays (denoted by tt(0$\ell$)\Hthadthad in the following),  as described in more detail in the next part of this section.   %
Most of these selections do not rely on the invariant mass of the two $\tau$-leptons, as the distribution of this variable is subsequently used in a likelihood fit to extract the signal as described in \cref{sec:stxs_results}. %
%

\begin{table}[tbhp]
\caption{
Summary of the inclusive signal region selection for the STXS measurement and the fiducial differential measurement. The $\pTH$ computation used in the Boost category is the same as in Ref.~\cite{HIGG-2019-09} and denoted by $\pTHreco$, and is not the NN-based \pTH  introduced in \cref{subsec:vbfcategorisation} for the VBF and ttH categories.   The leading and sub-leading jets are denoted by $j_{0}$ and $j_{1}$, respectively.
}
\label{tab:selection:signalregions}
\begin{center}
\scalebox{0.9}{
\begin{tabular}{r|c|c|c|c}
\hline
& \multicolumn{4}{c}{ } \\
\textbf{VBF inclusive} & \multicolumn{4}{c}{$N_{\text{jets}} \geq$ 2,  sub-leading jet $\pT > \SI{30}{\GeV}$} \\
\textbf{STXS}          & \multicolumn{4}{c}{$\mjj > \SI{350}{\GeV}$, $|\detajj| > 3$} \\
& \multicolumn{4}{c}{$\eta(j_{0}) \times \eta(j_{1}) < 0$} \\
& \multicolumn{4}{c}{lepton centrality: visible decay products of the $\tau$-leptons between VBF jets} \\
& \multicolumn{4}{c}{ } \\
\hline
& \multicolumn{4}{c}{ } \\
\textbf{VBF inclusive} & \multicolumn{4}{c}{$N_{\text{jets}} \geq$ 2,  sub-leading jet $\pT > \SI{30}{\GeV}$} \\
\textbf{differential}  & \multicolumn{4}{c}{$\mjj > \SI{600}{\GeV}$, $|\detajj| > 3.4$, $\ptjj > \SI{30}{\GeV}$} \\
& \multicolumn{4}{c}{$\eta(j_{0}) \times \eta(j_{1}) < 0$} \\
& \multicolumn{4}{c}{lepton centrality: visible decay products of the $\tau$-leptons between VBF jets} \\
& \multicolumn{4}{c}{$\pTHjj < \SI{50}{\GeV}$} \\
& \multicolumn{4}{c}{ } \\
\hline
& \multicolumn{4}{c}{ } \\
\textbf{VH inclusive} & \multicolumn{4}{c}{$N_{\text{jets}} \geq$ 2,  sub-leading jet $\pT > \SI{30}{\GeV}$} \\
& \multicolumn{4}{c}{$\SI{60}{\GeV} < \mjj < \SI{120}{\GeV}$} \\
& \multicolumn{4}{c}{ } \\
\hline
& \multicolumn{4}{c}{ } \\
\textbf{tt(0$\ell$)\Hthadthad}   & \multicolumn{4}{c}{ $N_{\text{jets}} \geq$ 6 and $N_{b\text{-jets}}$ $\geq$ 1 }\\
& \multicolumn{4}{c}{ or $N_{\text{jets}}\geq$ 5 and $N_{b\text{-jets}}$ $\geq$ 2} \\
& \multicolumn{4}{c}{ } \\
\hline
& \multicolumn{4}{c}{ } \\
\textbf{Boost inclusive} & \multicolumn{4}{c}{Not VBF inclusive} \\
& \multicolumn{4}{c}{Not VH inclusive} \\
& \multicolumn{4}{c}{$\pTHreco > \SI{100}{\GeV}$} \\
& \multicolumn{4}{c}{ } \\
\hline
\end{tabular}
}
\end{center}

\end{table}


\subsection{VBF categorization}
\label{subsec:vbfcategorisation}

The VBF inclusive region is designed to select events with a Higgs boson produced in association with two jets originating from the hard scattering.
These events can be distinguished from those from other Higgs boson production modes, and also from \Zttjets events, because they contain two high-\pT jets with a large rapidity gap and a large dijet invariant mass (\mjj).
In order to further reduce the ggF contamination for the differential fiducial measurement, requirements on \mjj and $|\detajj|$ are increased and additional requirements on the dijet transverse momentum (\ptjj) and the transverse momentum of the Higgs boson decay products and the two jets (\pTHjj) are introduced in the VBF inclusive region.

Following the convention of the STXS framework, events in the VBF inclusive region are split into eight bins according to $\mjj \in [350, 700, 1000, 1500, \infty)$~GeV and $\pTH\gtrless 200$~GeV.
A novel technique based on a neural network (NN) is adopted to estimate $\pTH$ for the VBF and tt(0$\ell$)\Hthadthad categories.
This NN is trained on ggF production events, using as input variables the $\Delta R$ and $\Delta\phi$ separation between the two $\tau$-leptons,  the $\pt$ of the system formed by the four-momenta of the two $\tau$-leptons and the \met ($\pTHreco$), and $m_{\tau\tau}^{\mathrm{coll}}$ information.
Since this NN exploits the invariant mass of the system and assumes $H\to\tau\tau$ decays,  it improves the $\pTH$ resolution in signal events -- by 50\% compared to that obtained with $\pTHreco$ alone -- but leads to misreconstruction of the $\pT$ for processes like $\Zttjets$.

A boosted decision tree (BDT) discriminant, trained using the dijet system's kinematic variables as described in \cref{tab:mva:variables}, is used to enhance the VBF $H\to\tau\tau$ signal relative to ggF $H\to\tau\tau$ production and the \Zttjets background.
This BDT was also used in Ref.~\cite{HIGG-2019-09}, where more details about it can be found,  but the selection criteria applied to the BDT score have been optimized in each of the eight VBF STXS bins and in the differential VBF inclusive region.
Based on the discriminant output, additional event splitting into two subcategories, VBF\_0 enriched in background and VBF\_1 enriched in signal events, is defined.

\begin{table}[tbh]
\caption{ Variables used in the two multivariate taggers employed in
the analysis.  For each tagger, the presence or absence of a $\bullet$
indicates whether the variable is used or not.
The symbol $\tau$ stands for any reconstructed $\tau$-lepton
candidate (electron, muon or $\tau_{\text{had-vis}}$) as appropriate in each
channel. The symbols \pttt and \ptjj indicate the vectorial sums of the momenta of
two visible $\tau$-lepton candidates and of the two leading jets, respectively.  The
Higgs boson candidate is formed by the vector sum of the two visible $\tau$-lepton
candidates' momenta and \vecmet.  The $W$-boson candidate in an event is built as the
pair of non-$b$-tagged jets with invariant mass closest
to $m_W$. The top-quark candidate in an event is built as the system of the $W$-boson
candidate and a $b$-tagged jet with invariant mass
closest to \mtop.}
\label{tab:mva:variables}
\centering
\begin{tabular}{@{}clcc@{}}
\toprule
& Variable                                           & VBF       & ttH multiclass \\
\midrule
& Invariant mass of the two leading jets                   & $\bullet$ &             \\
& \ptjj                                              & $\bullet$ &             \\
& Product of $\eta$ of the two leading jets                & $\bullet$ &             \\
& Sub-leading jet \pT                                 & $\bullet$ &             \\
Jet properties & $\eta$ of the 5 leading jets                                 &           & $\bullet$   \\
& Scalar sum of all jets \pT                         &           & $\bullet$   \\
& Scalar sum of all $b$-tagged jets \pT              &           & $\bullet$   \\
& Best $W$-boson candidate dijet invariant mass            &           & $\bullet$   \\
& Best top-quark candidate three-jet invariant mass &           & $\bullet$   \\
\midrule
& \dphi between the two leading jets                        & $\bullet$ &             \\
& \deta between the two leading jets   & $\bullet$ &             \\
Angular & Minimum \dr between two jets    &         & $\bullet$    \\
distances & Minimum \dr between a $b$-tagged jet and a $\tau_{\text{had-vis}}$ &         & $\bullet$    \\
& $|\deta(\tau,\tau)|$                               &           & $\bullet$   \\
& $\dr(\tau,\tau)$                               &           & $\bullet$   \\
\midrule
$\tau$-lepton & \pttt                                          &           & $\bullet$     \\
properties & Sub-leading $\tau$ \pT                              &           & $\bullet$   \\
& Leading $\tau$ $\eta$                           &           & $\bullet$   \\
\midrule
\multirow{1}{*}{$H$ candidate plus}
& \pTHjj                                             & $\bullet$ &             \\
\multirow{1}{*}{jets system} & \\
\midrule
\vecmet & Missing transverse energy \MET                     &           & $\bullet$   \\
& Smallest $\dphi(\tau, \vecmet$)                    &           & $\bullet$   \\
\bottomrule
\end{tabular}
\end{table}

\begin{figure}[ht]
\begin{center}
\begin{tabular}{c}
\includegraphics[width=\linewidth]{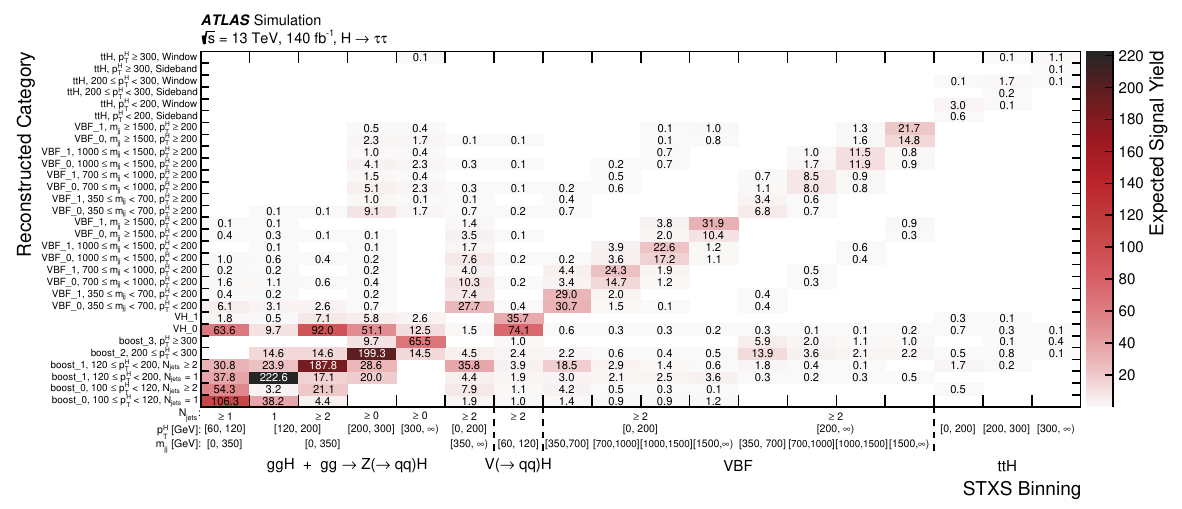} \\
(a) \\
\includegraphics[width=\linewidth]{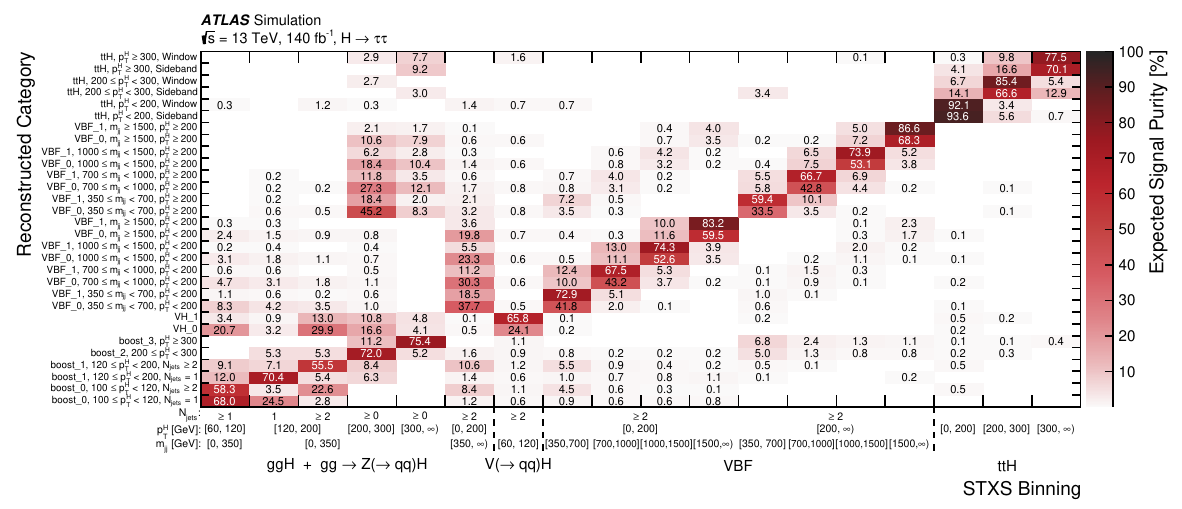} \\
(b) \\
\end{tabular}
\caption{
(a)~Expected SM $H\to\tau\tau$ signal yield and (b)~expected signal purity in each signal region of the analysis ($y$-axis).
Yields are summed over the three $\tau\tau$ channels ($\tau_{e}\tau_{\mu}$, $\tau_{\text{lep}}\tau_{\text{had}}$, $\tau_{\text{had}}\tau_{\text{had}}$).
Only entries with a value above 0.1 are shown in the plots.
\label{fig:signal:yields}}
\end{center}
\end{figure}

%
%
The threshold values for the BDT output are optimized to maximize the figure of merit  $\sigma_{\mathrm{tot}} = \sqrt{\sigma^2_0+\sigma^2_1}$, where $\sigma_i = S_i/\sqrt{S_i+B_i}$ with $S_i$ and $B_i$ being the numbers of signal and background events in the VBF\_0 ($i=0$) and VBF\_1  ($i=1$) categories.
This results in the VBF inclusive STXS region being split into a total of 16 signal regions. %
The signal yield expected in each of these signal regions is shown in \cref{fig:signal:yields}(a), while \cref{fig:signal:yields}(b) illustrates the relative signal population in each signal region.

The differential fiducial measurement has two signal regions:  VBF\_0 enriched in background and VBF\_1 enriched in signal events.

\subsection{tt(0$\ell$)\Hthadthad categorization}
\label{sec:sel_tth}

The tt(0$\ell$)\Hthadthad region is designed to enhance the contribution from $\ttH$ production. Events are required to have at least six jets including at least one $b$-tagged jet, or five jets including at least two $b$-tagged jets.
Events within this category are split into several regions based on a multiclass BDT  trained to discriminate the \ttH signal events from the \Ztt and $t\bar{t}$ background events. %
A detailed description of the kinematic variables used for the multiclass BDT training is given in \cref{tab:mva:variables}.
Regions enhanced in signal events are defined by combining requirements on the BDT scores; the threshold values for the scores  are optimized using the same figure of merit as the one used for VBF categorization. %
By inverting the requirement on the score, separate control regions for the \Ztt and $t\bar{t}$ backgrounds are also defined.
According to the prescription of the STXS framework, three signal regions enriched in signal events are defined according to $\pTH \in [0, 200, 300, \infty)$~GeV. The NN-based \pTHNN described in \cref{subsec:vbfcategorisation} is used to make this splitting.
Each signal region is also required to have $m_{\tau\tau} \in [100, 150]$~GeV, while events outside this window are assigned to a sideband region that is used to control the backgrounds.
The signal yield expected in each of these regions is shown in \cref{fig:signal:yields}(a), while \cref{fig:signal:yields}(b) illustrates the relative signal population in each signal region.

\subsection{V(had)H categorization}

To match the STXS $qq\to V(\to qq)H$ particle-level \pTjet requirement and \mjj binning, events selected in the V(had) categories must satisfy $\SI{60}{\GeV}<\mjj<\SI{120}{\GeV}$ and the \pT of the sub-leading jet must be greater than \SI{30}{\GeV}.
Events in this category are split into two subcategories using a dedicated tagger.
The V(had)H tagger was trained by treating all Higgs boson events produced by processes other than $VH$ as background.
The BDT score requirement used to define the two categories was optimized to give the smallest uncertainty for the V(had)H cross-section, and provides a selection where the expected fraction of V(had)H among all Higgs boson events is \SI{66}{\percent} (\SI{24}{\percent}) in the VH\_1 (VH\_0) category.
The signal yield expected in each of these regions is shown in \cref{fig:signal:yields}(a), while \cref{fig:signal:yields}(b) illustrates the relative signal population in each signal region.

\subsection{Boost categorization}

Events failing to meet the criteria of the VBF, V(had)H and ttH categories but having a high-\pT Higgs boson candidate are considered for the `Boost' categories targeting ggF events with large Higgs boson transverse momentum.
Events in the Boost category must satisfy $\pTH>\SI{100}{\GeV}$.
The reconstructed Higgs boson's transverse momentum, \pTH, is estimated using $\pTHreco$ since no significant improvement was found in this Boost category by using the NN-based $\pTH$ reconstruction.
Following the convention of the STXS framework, events are further categorized by \pTH value and by the total number of jets with \pT greater than \SI{30}{\GeV} (\njetsthirty). %
Events with $\pTH<\SI{200}{\GeV}$ are separated into 1-jet and $\ge$2-jet categories, while events with $\pTH>\SI{200}{\GeV}$ and at least one jet are considered without further categorization based on jet multiplicity.
The signal yield expected in each of these regions is shown in \cref{fig:signal:yields}(a), while \cref{fig:signal:yields}(b) illustrates the relative signal population in each signal region.

%
%
%
%
%
%
%
%
%
%
%
%
%
%
%
%
%

%
%
%

%
%
%
%
%
%
%
%
%
%
%
%
%
%
%


%
%
%
%

%
\section{Background estimation }
\label{sec:background}

The dominant backgrounds after the event selection are irreducible backgrounds from \Zttjets and \ttbar production, and reducible backgrounds from processes where at least one jet is misidentified as a \tauhad or a light lepton.
The contributions from these processes are evaluated using a mixture of simulations and data-driven techniques, following the strategy described extensively in Ref.~\cite{HIGG-2019-09}.

Events from the \Zttjets process form the dominant background and they are estimated using MC simulated samples constrained using data.
The yields predicted from  \Zttjets MC samples are corrected using dedicated control regions defined by selecting events with kinematic properties similar to those in the corresponding signal regions,  as described in detail in Ref.~\cite{HIGG-2019-09}.
The selection used to define these control regions starts from \Zlljets data and MC events which are modified through a simplified implementation of the \emph{embedding} procedure~\cite{HIGG-2014-09}, rescaling the transverse momentum of each reconstructed lepton through parameterizations to account for the differences in the trigger efficiencies, the kinematics of prompt leptons and the decay products of $\tau$-leptons. After the rescaling, all the relevant kinematic quantities in the analysis are recomputed. This also includes accounting for the additional \met{} from the neutrinos from the decay of the $\tau$-leptons.
These control regions are included in the fit to help constrain the normalization of \Zttjets MC events in the signal regions.
The uncertainties affecting the reconstructed objects used in the embedding are propagated through the entire procedure, and include those associated with the parameterizations.
Dedicated uncertainties affecting each control region are assigned to account for the modelling differences observed between the \Ztt and embedded \Zll MC predictions because of the approximations associated with the simplified embedding procedure. For the fiducial differential measurement an additional shape modelling uncertainty is assigned to the variable under study by comparing the embedded data and MC samples for each variable in each control region.

Processes with at least one jet misidentified as an electron, muon or \tauhad are collectively referred to as misidentified $\tau$ background.
They are evaluated using data-driven techniques in the three channels considered in the analysis, following the same methodology as in Ref.~\cite{HIGG-2019-09}.

In the \temu channel, the misidentified $\tau$ background is estimated using the matrix method~\cite{EGAM-2019-01}.
Data events are selected using the same criteria as in \cref{tab:selection:baseline}, but without the lepton isolation criteria and with loosened identification criteria for the electrons.
The expected number of fake leptons in each signal region is computed from a system of equations relating the efficiencies for real and fake leptons to the observed event yields.
The real-lepton efficiencies are estimated using simulations corrected with data-based scale factors, while the fake-lepton efficiencies are measured using data events which have two leptons of the same charge.
Uncertainties are assigned to this background, taking into account the statistical uncertainties in the derived efficiencies, the dependency on the number of jets in the final state, the different MC samples used to derive the efficiencies, and the normalization of the real-lepton component subtracted in the efficiency estimation.

In the \tlhad channel, the misidentified $\tau$ background is evaluated using the fake-factor technique~\cite{EGAM-2019-01}.
Data events are selected if they satisfy a very loose requirement on the \tauhadvis identification score but do not satisfy the \objectsc{Medium} working point criteria.
Residual contributions from processes with real \tauhadvis satisfying this requirement are evaluated using simulations and subtracted.
The distribution of the misidentified $\tau$ background component in each signal region is obtained by multiplying the contribution of data events selected with the inverted identification criterion by a fake-factor defined as the ratio of misidentified \tauhadvis that respectively pass or fail the \objectsc{Medium} working point of the \tauhadvis identification algorithm.
Uncertainties are assigned to this background, taking into account the statistical uncertainties in the fake-factors, the uncertainties associated with the subtraction of the residual contributions from processes with real \tauhadvis, and the uncertainties in the flavour composition estimate.

In the \thadhad channel, the misidentified $\tau$ background is also determined using a fake-factor approach.
The method is similar to the one used in the \tlhad channel, and it takes into account the processes with one or two jets misidentified as \tauhadvis and the trigger matching.
Uncertainties are assigned to this background, taking into account the statistical uncertainties in the fake-factors, differences between fake-factors calculated in different validation regions, and different fake-factor parameterizations.

Other minor backgrounds arising from an electron or a muon misidentified as a \tauhad are estimated from simulation and, in the case of electrons, corrected for the efficiency measured in data.

Event contributions from the \ttbar process are sizeable in the \temu and \tlhad channels, and control regions enriched in top-quark events are defined by replacing the $b$-jet veto in the event selection (see \cref{tab:selection:baseline}) with a requirement of at least one $b$-tagged jet. This allows the \ttbar MC simulation to be validated, and the \ttbar process normalization to be determined using data.


%
\section{Systematic uncertainties}
\label{sec:systematic}

Systematic uncertainties affect the yields in the various signal and control regions as well as the distribution shape of the main fit observable ($m_{\tau\tau}$).
They can be assigned to three main groups: the experimental uncertainties, the theoretical uncertainties for the backgrounds and the theoretical uncertainties for the signal.
Overall, the systematic uncertainties having largest impact on the measurement are similar to those in Ref.~\cite{HIGG-2019-09}.

The experimental systematic uncertainties include those affecting the trigger, reconstruction, identification and isolation efficiencies for all the considered particles in the final state, as well as their energy scale and resolution~\cite{ATLAS:2023dxj,MUON-2018-03, ATLAS-CONF-2017-029,JETM-2018-05,PERF-2014-03,FTAG-2018-01,FTAG-2020-08,FTAG-2019-02}.
These uncertainties also affect the shape of the $m_{\tau\tau}$ distribution, the background yields and the signal cross-section through their effects on the acceptance and the migration between different event categories.
Among the experimental uncertainties, the jet-related uncertainties in the energy scale and resolution are the ones having largest impact on the results discussed in Section~\ref{sec:stxs_results}.
The jet energy scale uncertainty includes components related to the in~situ calibration of jets as well as pile-up, the extrapolation to higher transverse momentum, and uncertainties in the different responses to quark- and gluon-initiated jets.
The jet energy scale uncertainty for central jets ($|\eta|<1.2$) varies from \SI{1}{\percent} for a wide range of jet \pt ($\SI{250}{\GeV}<\pt<\SI{2000}{\GeV}$) to \SI{5}{\percent} for very low \pt jets (\SI{20}{\GeV}) and to \SI{3.5}{\percent} for very high \pt jets ($>\SI{2.5}{\TeV}$).
The jet energy resolution is measured in a dedicated analysis~\cite{JETM-2018-05} and ranges from ($24\pm5$)\% at \SI{20}{\GeV} to ($6.0\pm0.5$)\% at \SI{300}{\GeV}.

The uncertainties in the \tauhadvis identification efficiency range from \SI{2}{\percent} to \SI{6}{\percent} and are parameterized as a function of the \tauhadvis \pt\ and number of associated tracks. The  eBDT efficiency uncertainties are of the order of \SI{1}{\percent} to \SI{2}{\percent} and are parameterized as a function of the \tauhadvis \pt, \tauhadvis $\eta$ and $\tau$-lepton decay mode.
Since the analysis is sensitive to the \tauhadvis reconstruction efficiency uncertainty through the usage of the kinematic embedding, this efficiency is left as a free parameter in the fit and measured in situ. %
The total uncertainty  of the \tauhadvis energy scale ranges from \SI{1}{\percent} to \SI{4}{\percent}, and is estimated by combining a direct measurement with \Ztaumutauhad events, measurements of the calorimeter response to single particles, and comparisons between simulations using different detector geometries or $\GEANT$ physics lists.
This uncertainty is also parameterized as a function of the \tauhadvis \pt\ and number of associated tracks~\cite{ATLAS-CONF-2017-029}.

For the theoretical uncertainties in the background, the largest impact on the results is due to the uncertainties associated with the \Zjets and  \ttbar samples.
For \Zjets, uncertainties were considered for the renormalization scale (\muR), factorization scale (\muF) and resummation scale (\muQ), for the jet-to-parton matching scheme (\ckkw), for the choice of $\alphas$ value, and for the choice of PDFs.
The effects of uncertainties in \muR and \muF due to missing higher orders and non-perturbative effects were evaluated using six scale variations in the matrix elements by factors of $0.5$ and $2$, excluding the extreme variations in opposite directions~\cite{Bothmann:2016nao}.
The effects of \muQ and \ckkw uncertainties were estimated from a comparison with alternative samples generated with a varied resummation scale or jet-to-parton matching scheme, respectively, while the effects of other uncertainties were evaluated from internal-weight variations in simulated samples.
For \ttbar, uncertainties were considered for the choice of matrix element and parton shower generators, the choice of model for initial- and final-state radiation (ISR and FSR respectively), and the choice of PDFs.
The effects of PDF uncertainties were estimated using the \PDFforLHC[15nlo] set of eigenvectors, while the other uncertainties were obtained from a comparison with alternative samples generated using the same settings as the default sample apart from the variation of the parameter for which the uncertainty is estimated.

Several sources of theoretical uncertainty in the signal are considered.
For all production modes, uncertainties are considered for the PDF and \alphas, the parton shower and hadronization model, and missing higher orders in the matrix element calculation.
The effects of PDF and \alphas uncertainties were estimated from the \PDFforLHC[15nlo] set of eigenvectors.
The impact of using a different parton shower and hadronization model is evaluated by comparing the nominal sample with an event sample generated using \HERWIG[7]~\cite{Bellm:2017jjp} instead of \PYTHIA[8].
The impact of using a different matrix element generator for the signal sample is evaluated by comparing the nominal signal sample with an alternative sample generated with \MGNLO~\cite{Alwall:2014hca} instead of \POWHEGBOX, while keeping the same parton shower model.

For the ggF production process, 18 sources of uncertainty related to the ggF kinematics and the cross-section were considered, following the prescription in Ref.~\cite{ATL-PHYS-PUB-2023-031}.
Two uncertainties are assigned for the overall fixed-order and resummation effects.
Two uncertainties cover the migrations between different jet-multiplicity bins.
Seven uncertainties are associated with the modelling the Higgs boson \pt  in different phase-space regions.
Four uncertainties take into account dijet mass migrations across the STXS bin boundaries.
One uncertainty covers the modelling of the Higgs boson plus two leading jets transverse momentum ($p_\text{T}^{Hjj}$) distribution in the $\ge$2-jet region.
One uncertainty is assigned to the modelling of the distribution of the Higgs boson plus one jet transverse momentum ($p_\text{T}^{Hj}$) divided by \pTH in the high-\pTH region. The last uncertainty takes into account the uncertainty from the choice of top-quark mass scheme.

For the VBF and $VH$ processes, ten uncertainties related to the STXS categorization were considered: one related to the inclusive cross-section of the process, one related to the two-jet requirement, one related to the Higgs boson \pt selection at \SI{200}{\GeV}, one related to the \pt balance between the Higgs boson and the dijet system in events with two or three jets, and six which take into account dijet mass migrations across the STXS bin boundaries.

For the \ttH process, six other uncertainties are included: one related to the inclusive cross-section of the process, and five migration uncertainties related to Higgs boson \pt boundaries in the STXS scheme.


\section{Simplified template cross-section measurement results}
\label{sec:stxs_results}

The statistical analysis for the measurements in this paper is based on a likelihood function constructed as the product of Poisson
probability terms over the bins of the input distributions, as shown in Eq.~(\ref{eqn:llh}):

\begin{equation}
\label{eqn:llh}
L\left(\vec{\alpha}, \vec{\theta}, \text{data}\right) = \prod_{k\in\text{cat}}
\prod_{b\in\text{bins}}   P(n_{k,b} | n^{\text{signal}}_{k,b}(\vec{\alpha}, \vec{\theta}) +
n^{\text{bkg}}_{k,b}(\vec{\theta})) \prod_{\theta\in\vec{\theta}} G(\theta),
\end{equation}

where $n_{k,b}$, $n^{\text{signal}}_{k,b}$ and $n^{\text{bkg}}_{k,b}$
stand for the number of observed events, the number of expected signal events and the
number of expected background events in bin $b$ of analysis category $k$, respectively.
The systematic uncertainties (described in Section~\ref{sec:systematic}) are incorporated in the likelihood function through nuisance parameters, $\vec{\theta}$, that are constrained by Gaussian probability terms, $G$, that multiply the Poisson probability terms, $P$.
The parameters of interest (POIs) of the model are  indicated with  $\vec{\alpha}$ and are estimated by maximizing the likelihood.
The test statistic is constructed from the profile likelihood ratio, and the confidence intervals for the parameters of interest are
derived using the asymptotic approximation~\cite{Cowan:2010js}.

\subsection{Simplified template cross-section measurement fit model}
For the STXS measurement, the likelihood function is built with the following signal regions (SRs) and control regions (CRs):
\begin{itemize}
\item Six SRs for the Boost category in each channel; the SRs are defined starting from the Boost inclusive category defined in \cref{tab:selection:signalregions} and further split according to the number of jets and $\pTH$. %
For each SR and channel, a corresponding \Zttjets CR is defined using the kinematic embedding described in \cref{sec:background}.
Moreover, a CR enriched in the \ttbar process is defined for each inclusive category in the \temu and \tlhad channels.
Summing over the SRs, \Zttjets CRs and \ttbar CRs, and considering the three channels, this leads to 38 Boost categories.

\item Two SRs for the VH category in each channel; the SRs are defined starting from the VH inclusive category  defined in \cref{tab:selection:signalregions} and further split into regions of high and low scores of a BDT trained to distinguish the VH production mode from the other Higgs boson production modes.
For each SR and channel, a corresponding \Zttjets CR is defined using the kinematic embedding.
Moreover, a CR enriched in the \ttbar process is defined for each inclusive category in the \temu and \tlhad channels.
Summing over the SRs, \Zttjets CRs and \ttbar CRs, and considering the three channels, 14 VH categories are included in the fit.

\item Sixteen SRs for the VBF category in each channel, as described in \cref{subsec:vbfcategorisation}.
For each SR and channel, a corresponding \Zttjets CR is defined using the kinematic embedding.
Moreover, a CR enriched in the \ttbar process is defined for each inclusive category in the \temu and \tlhad channels.
Summing over the SRs, \Zttjets CRs and \ttbar CRs, and considering the three channels, this leads to 98 VBF categories.

\item Six SRs and two CRs for the ttH category in the \thadhad channel, as described in \cref{sec:sel_tth}.
\end{itemize}

In the SRs, the binned $m_{\tau\tau}$ distribution is used to obtain information about the cross-section relative to the SM prediction. In the Boost and VH SRs, $m_{\tau\tau}$ is reconstructed using only the MMC technique, while in the VBF and ttH SRs $m_{\tau\tau}^{\mathrm{coll}}$ is used for events where the  MMC technique failed to estimate the invariant mass. CRs are employed to constrain the event yield of specific background processes with a single-bin histogram containing the number of events in the corresponding CR. In particular, the \Zttjets CR provides additional constraining power for the \Zttjets background uncertainties, being complementary to the mass distribution in a region centred around the $Z$ boson peak.
Three different measurements are performed:

\begin{enumerate}
\item \emph{$pp\to\Htautau$ total cross-section}:
a single POI, corresponding to the $pp\to\Htautau$ cross-section, is estimated by the fit. In the likelihood function, the signal yields in each category are parameterized as the product of the $pp\to\Htautau$ cross-section, the integrated luminosity and the efficiency (including the acceptance of the ATLAS detector) of the selection for a SM Higgs boson with a mass of \SI{125.09}{\GeV}.
In this measurement, the relative contributions to the $pp\to\Htautau$ cross-section from the various production modes are fixed to the SM predictions.%

\item \emph{Cross-section per production mode}:
four POIs, corresponding to the cross-sections of the four dominant production modes (ggF, VBF, $VH$, \ttH) of the Higgs boson, are estimated by the fit. In this configuration, the event yields in the likelihood function are the sum of those from each individual production mode, parameterized as a function of the POI similarly to the way for the total cross-section measurement described above.

\item \emph{Simplified Template Cross-Section}:
eighteen POIs, corresponding to the cross-sections for different bins of the STXS \emph{stage 1.2} framework, are determined by the fit.
The cross-sections are measured for six ggF bins, three \ttH bins and nine $qq'\to H qq'$ bins (comprising \text{VBF} and ${qq'}\to{V}(\to{qq'})H$ production).
These bins were chosen by combining some of the STXS bins proposed in Ref.~\cite{deFlorian:2016spz}, taking into account the available signal sample size and the expected sensitivity of the analysis.
The ${qq}\to{V}(\to{qq})H$ contribution is measured for events with particle-level dijet mass \mjj between \SI{60}{\GeV} and \SI{120}{\GeV}, while the $\text{VBF}$ contribution is measured in events with \mjj above \SI{350}{\GeV}.
\end{enumerate}

\subsection{Inclusive and simplified template cross-section results}

\begin{figure}[htb!]
\begin{center}
\begin{tabular}{cc}
\includegraphics[width=0.4\linewidth]{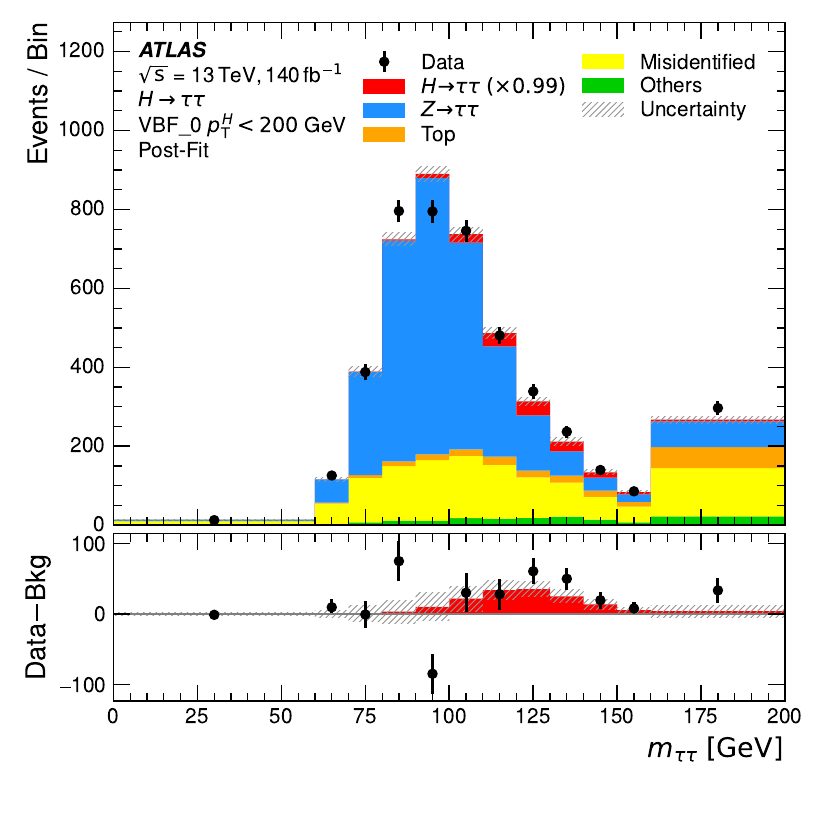} &
\includegraphics[width=0.4\linewidth]{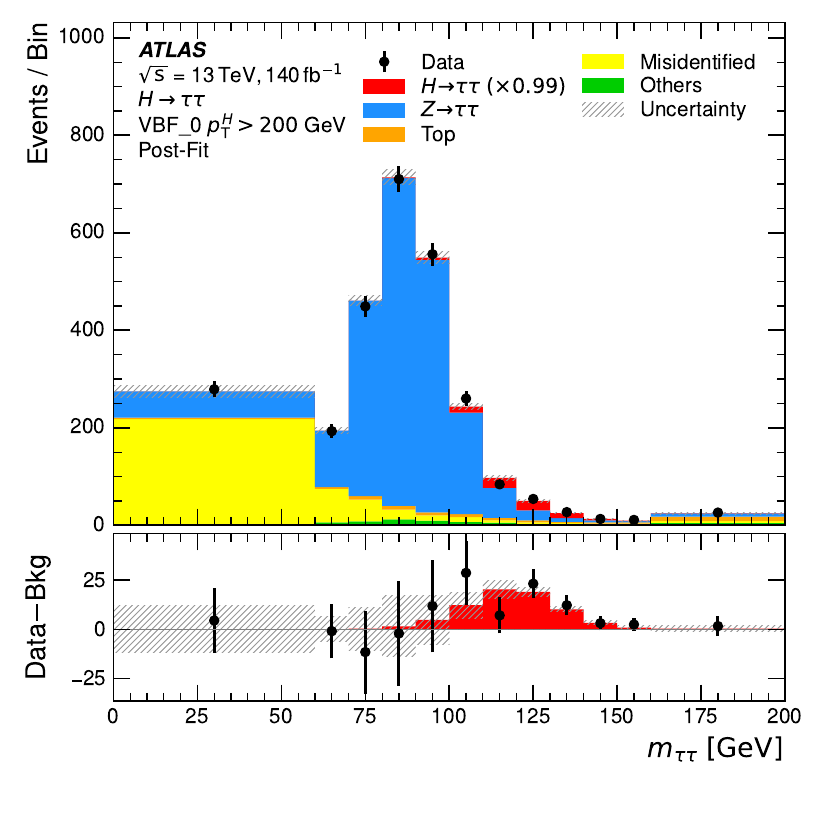} \\
(a) & (b) \\
\includegraphics[width=0.4\linewidth]{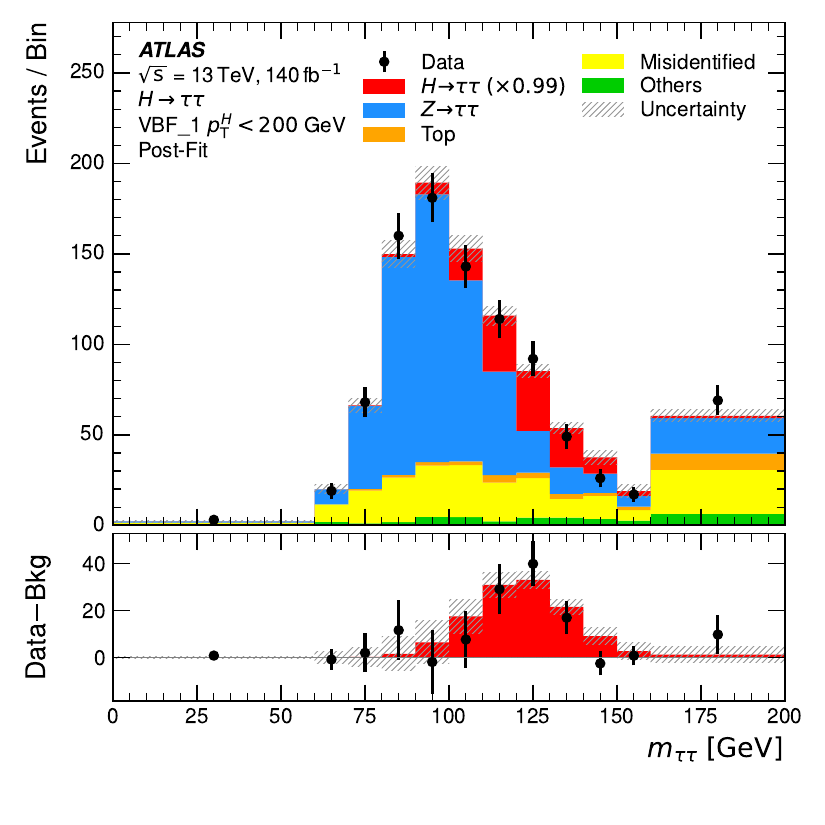} &
\includegraphics[width=0.4\linewidth]{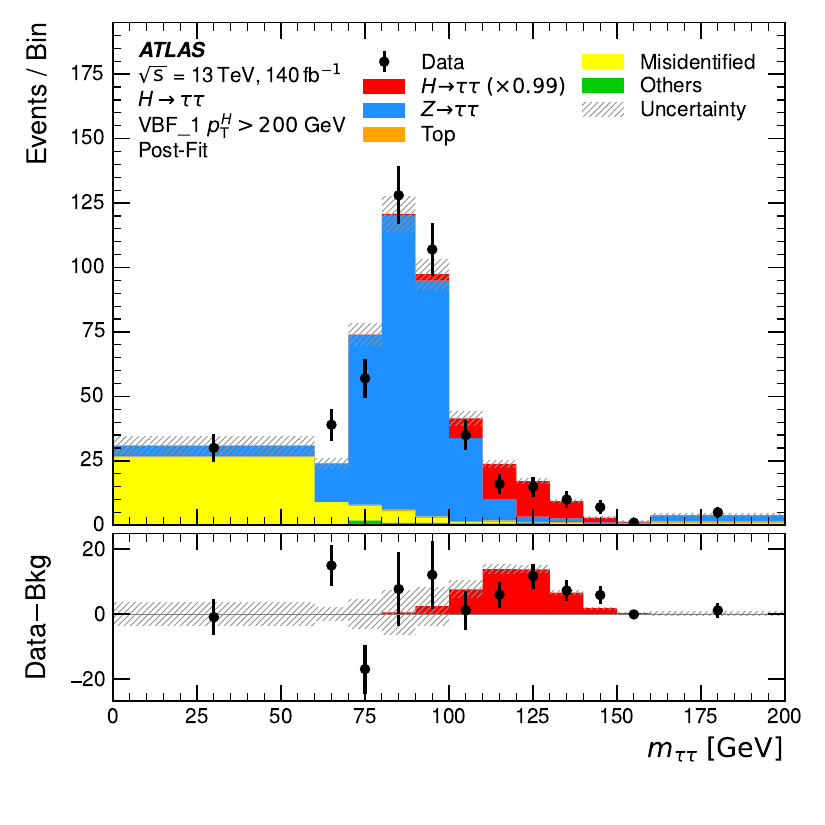} \\
(c) & (d) \\
\end{tabular}
\caption{
Distribution of the reconstructed $\tau\tau$ invariant mass ($m_{\tau\tau}$) for
all events in the (a,b)  VBF\_0 and (c,d)  VBF\_1 signal
regions for (a,c) $\pTH<200$~GeV and (b,d) $\pTH>200$~GeV.
The bottom panel shows the differences between observed data
events and expected background events (black points).  The observed
Higgs boson signal, corresponding to $(\sigma\times B)/(\sigma\times B)_{\text{SM}}\,=\,0.99$, is shown
with a filled red histogram.  Entries with values above the $x$-axis
range are shown in the last bin of each distribution.  The prediction
for each sample is determined from the likelihood fit performed to
measure the total $pp\,\to\Htautau$ cross-section.
\label{fig:stxs:low_high_pth}}
\end{center}
\end{figure}

\begin{figure}[htb!]
\begin{center}
\begin{tabular}{cc}
\includegraphics[width=0.4\linewidth]{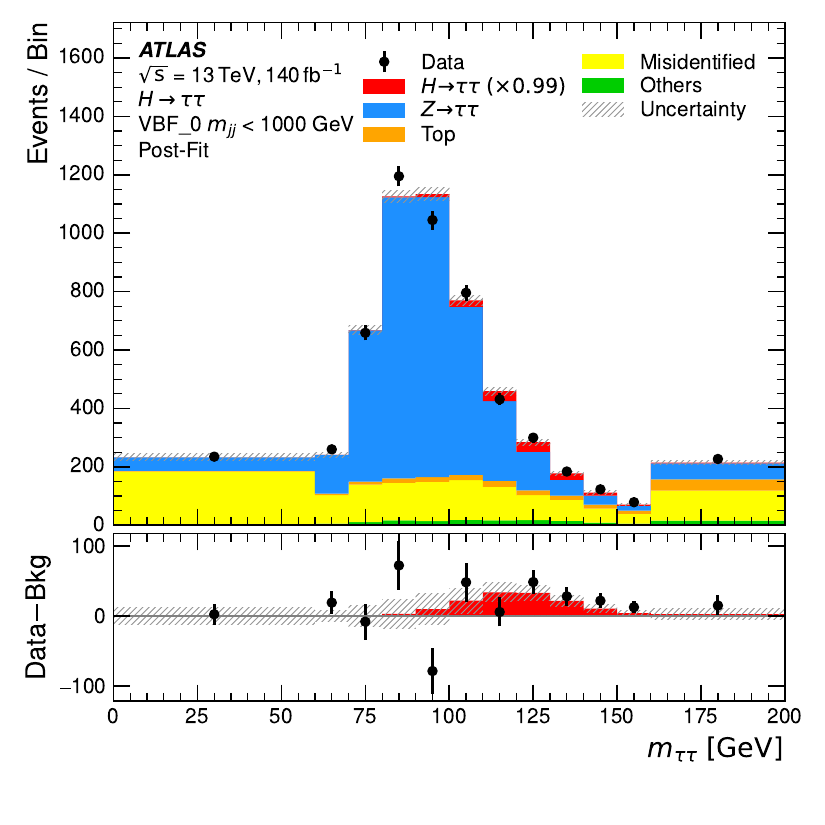} &
\includegraphics[width=0.4\linewidth]{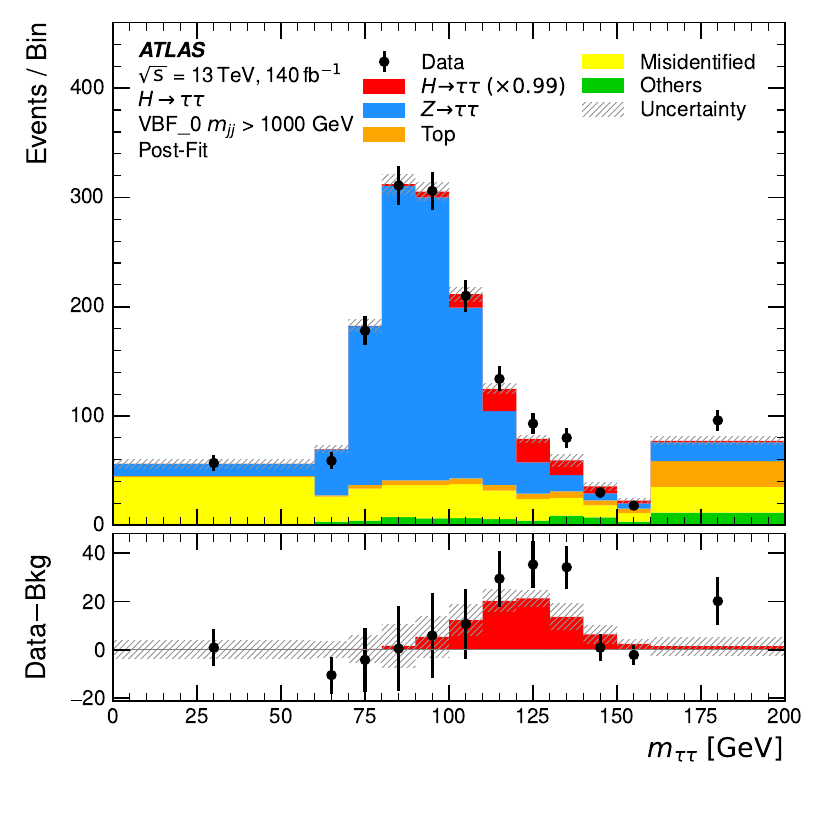} \\
(a) & (b) \\
\includegraphics[width=0.4\linewidth]{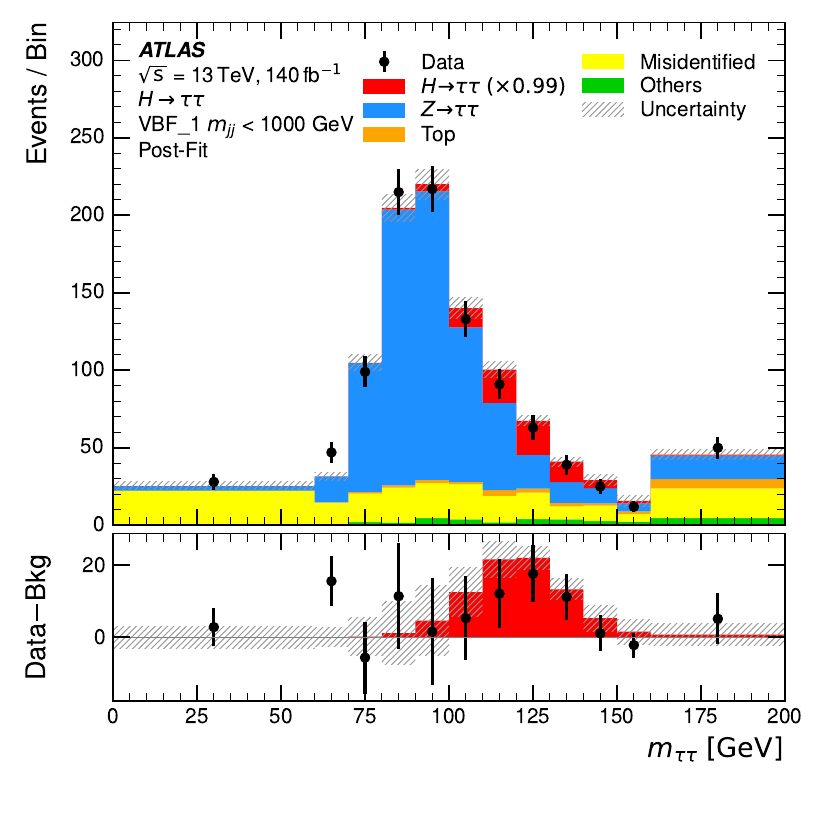} &
\includegraphics[width=0.4\linewidth]{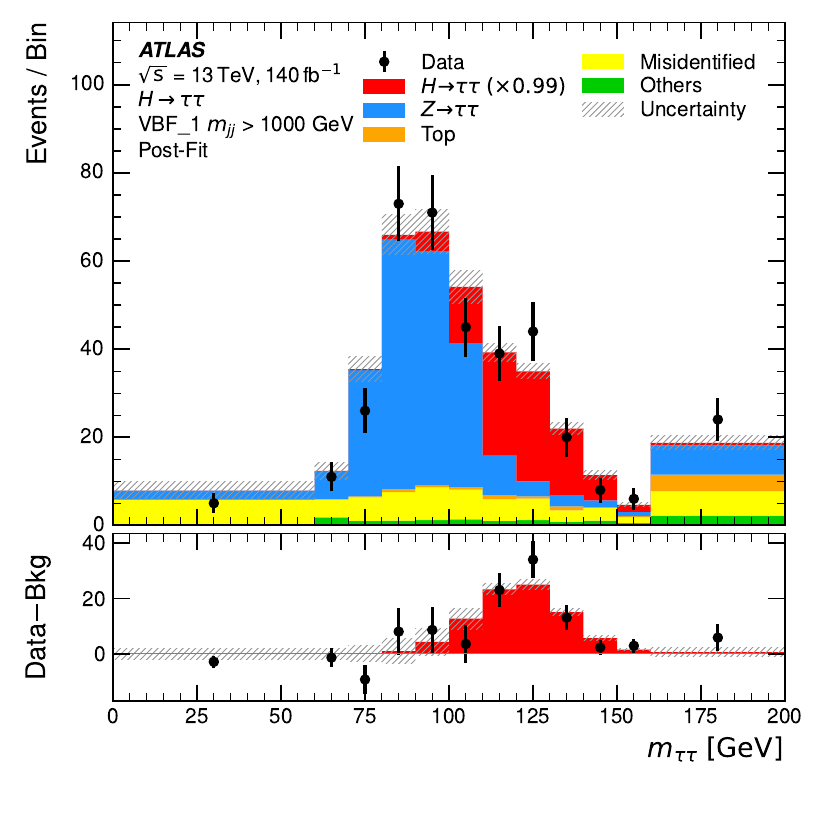} \\
(c) & (d) \\
\end{tabular}
\caption{
Distribution of the reconstructed $\tau\tau$ invariant mass ($m_{\tau\tau}$) for
all events in the (a,b) VBF\_0 and (c,d) VBF\_1 signal
regions for (a,c) $\mjj<1$~TeV and (b,d) $\mjj>1$~TeV.
The bottom panel shows the differences between observed data
events and expected background events (black points).  The observed
Higgs boson signal, corresponding to $(\sigma\times B)/(\sigma\times B)_{\text{SM}}\,=\,0.99$, is shown
with a filled red histogram.  Entries with values above the $x$-axis
range are shown in the last bin of each distribution.  The prediction
for each sample is determined from the likelihood fit performed to
measure the total $pp\,\to\Htautau$ cross-section.
\label{fig:stxs:low_high_mjj}}
\end{center}
\end{figure}

\begin{figure}[htb!]
\begin{center}
\includegraphics[width=0.8\linewidth]{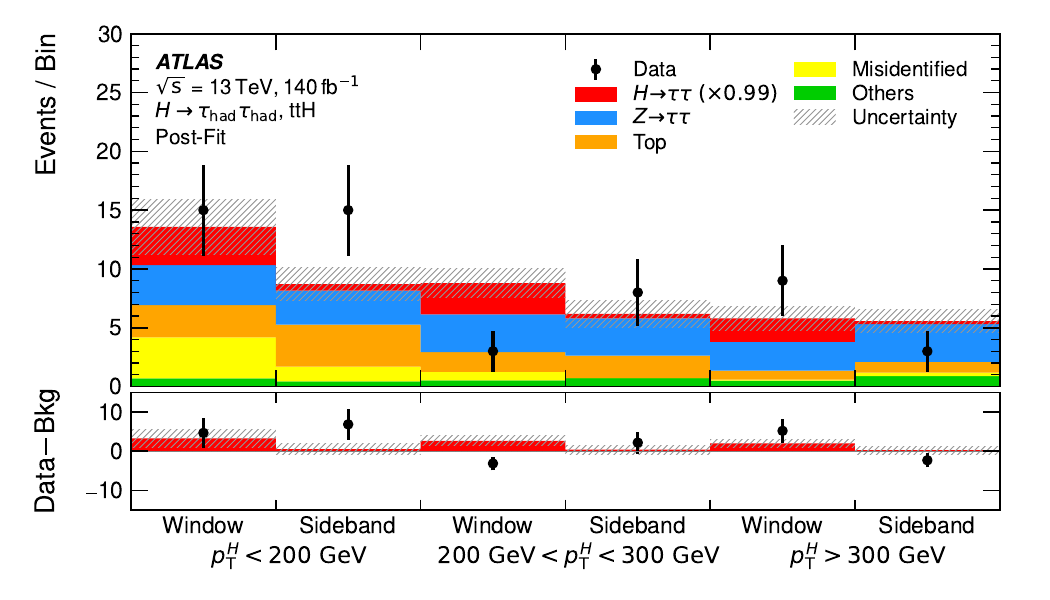}
\caption{
Comparison of the observed and expected event yields in the $\ttH$ categories.
The \emph{window} category contains events with $m_{\tau\tau} \in [100, 150]$~GeV, while events outside of this mass region are included in the \emph{sideband} region.
The bottom panel shows the differences between observed data
events and expected background events (black points).  The observed
Higgs boson signal, corresponding to $(\sigma\times B)/(\sigma\times B)_{\text{SM}}\,=\,0.99$, is shown
with a filled red histogram. The prediction
for each sample is determined from the likelihood fit performed to
measure the total $pp\,\to\Htautau$ cross-section.
\label{fig:stxs:tth_summary}}
\end{center}
\end{figure}

The results of the statistical analysis performed for the total cross-section measurement are presented in \cref{fig:stxs:low_high_pth,fig:stxs:low_high_mjj,fig:stxs:tth_summary}.
\cref{fig:stxs:low_high_pth,fig:stxs:low_high_mjj} show the $\tau\tau$ invariant mass distributions, split into low and high ranges of $\pTH$ and $\mjj$ respectively, in VBF\_0 and VBF\_1 for a subset of the categories used in the analysis. The signal from $\Htautau$ is clearly visible on top of the background originating mostly from $\Ztt$.
\cref{fig:stxs:tth_summary} shows the expected and observed event yields in the \ttH categories.

\begin{figure}[htb!]
\begin{center}
\includegraphics[width=0.7\linewidth]{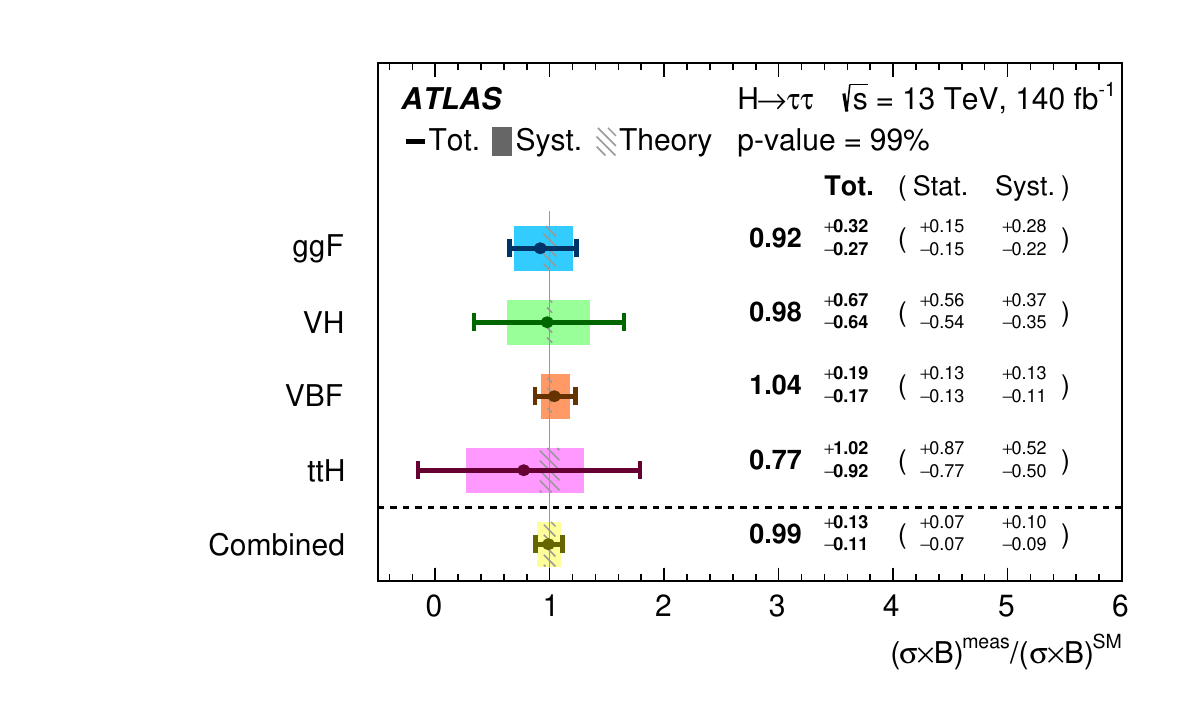}
\caption{
The measured values of $\sigma_{H}\times B$(\Htautau) relative to the SM
expectations in the total (labelled as `Combined') and per-production-mode fit.  The total $\pm1\sigma$
uncertainty in each measurement is indicated by the error bar,
with the contribution from the systematic uncertainty indicated by the coloured
band.}
\label{fig:stxs:4pois}
\end{center}
\end{figure}

The measurements of the total cross-section and per-production-mode cross-sections are shown in \cref{fig:stxs:4pois}. %
\cref{tab:grouped_impacts} shows the impact of the different uncertainty sources on the measurements, while the cross-section measured for each production mode is reported in \cref{tab:cross_section_4poi_tab}.
Good agreement with the SM predictions is found, with $p$-values of 97\% and 99\% for compatibility of the measurements with the SM for the total cross-section and per-production-mode cross-sections, respectively.
The VBF production cross-section normalized to the SM prediction is measured to be $1.04^{+0.19}_{-0.17}$, with slightly better precision than in Ref.~\cite{HIGG-2019-09} ($0.90^{+0.20}_{-0.17}$) due to the improved categorization used in the present analysis.
The $\ttH$ cross-section normalized to the SM prediction is measured to be $0.77^{+1.02}_{-0.92}$, using a more refined multivariate analysis strategy which improves the expected precision by roughly 25$\%$ relative to the result in Ref.~\cite{HIGG-2019-09} ($1.06^{+1.28}_{-1.08}$).

\begin{table}[h]
\centering
\caption{Summary of the different sources of uncertainty in the ratio of $\sigma_{H}\times B$(\Htautau) to the SM expectation for each production mode.
Experimental uncertainties for reconstructed objects combine efficiency and energy/momentum scale and resolution uncertainties.
Samples size includes the bin-by-bin statistical uncertainties in the simulated backgrounds as well as statistical uncertainties in misidentified $\tau$ backgrounds, which are estimated using data.
Entries with no significant impact on the measurement are denoted by `\textendash'.
}
\label{tab:grouped_impacts}
\begin{tabular}{ccccc}
\toprule
Production mode  & ggF         & ttH         & VBF         & VH          \\
\midrule
Best-fit value                                     & ~~\,0.92    & ~~\,0.77   & ~~\,1.04    & ~~\,0.98        \\
Total uncertainty                                  & $\pm 0.30$  & $\pm 0.97$ & $\pm 0.18$  & $\pm 0.66$ \\
\midrule
Statistical uncertainty                            & $\pm 0.15$  & $\pm 0.82$ & $\pm 0.13$ & $\pm 0.55$  \\
Total systematic uncertainty                       & $\pm 0.25$ & $\pm 0.51$ & $\pm 0.12$ & $\pm 0.36$ \\
\midrule
Samples size                                       & $\pm 0.09$ & $\pm 0.31$ & $\pm 0.04$ & $\pm 0.26$ \\
Theoretical uncertainty in signal                  & $\pm 0.18$ & $\pm 0.12$ & $\pm 0.11$ & $\pm 0.14$ \\
Jet and \etmiss                                    & $\pm 0.12$ & $\pm 0.15$ & $\pm 0.03$ & $\pm 0.12$ \\
Hadronic $\tau$-lepton decays                      & $\pm 0.04$ & $\pm 0.09$ & $\pm 0.01$ & $\pm 0.04$ \\
Misidentified $\tau$-lepton background             & $\pm 0.04$ & $\pm 0.05$ & $\pm 0.02$ & $\pm 0.11$  \\
Luminosity                                         & $\pm 0.01$ & ~~$\pm 0.01$ & $\pm 0.01$ & $\pm 0.01$ \\
Theoretical uncertainty in top-quark processes     & $\pm 0.01$ & $\pm 0.31$ & \textendash & $\pm 0.02$ \\
Theoretical uncertainty in \Zjets\ processes       & $\pm 0.02$ & $\pm 0.08$ & \textendash & $\pm 0.02$ \\
Flavour tagging                                    & $\pm 0.02$ & $\pm 0.05$ & $\pm 0.01$ & $\pm 0.01$ \\
Electrons and muons                                & $\pm 0.02$ & $\pm 0.02$ & $\pm 0.01$ & $\pm 0.02$ \\
\bottomrule
\end{tabular}
\end{table}

\begin{table}[h]
\centering
\caption{
Best-fit values and uncertainties of the $\sigma_{H}\times B$(\Htautau) for the four dominant production modes. The uncertainties include both the statistical and systematic components.
The SM predictions for each region, computed using the inclusive cross-section calculations and the simulated event samples, are also shown.
}
\begin{tabular}{r|c|c}
\toprule
Production mode & \multicolumn{2}{c}{$\sigma_{H}\times B$(\Htautau) [pb]} \\

& SM prediction & Measurement \\
\midrule
ggF          &  2.77  $\pm$  0.09   & 2.5   $\pm$ 0.8   \\
$VH$         &  0.117 $\pm$ 0.003 & 0.11  $\pm$ 0.08  \\
VBF          &  0.220  $\pm$ 0.005  & 0.23  $\pm$ 0.04  \\
\ttH         &  0.031 $\pm$ 0.003 & 0.02  $\pm$ 0.03  \\
\bottomrule
\end{tabular}
\label{tab:cross_section_4poi_tab}
\end{table}

\begin{figure}[htb!]
\begin{center}
\includegraphics[width=0.9\linewidth]{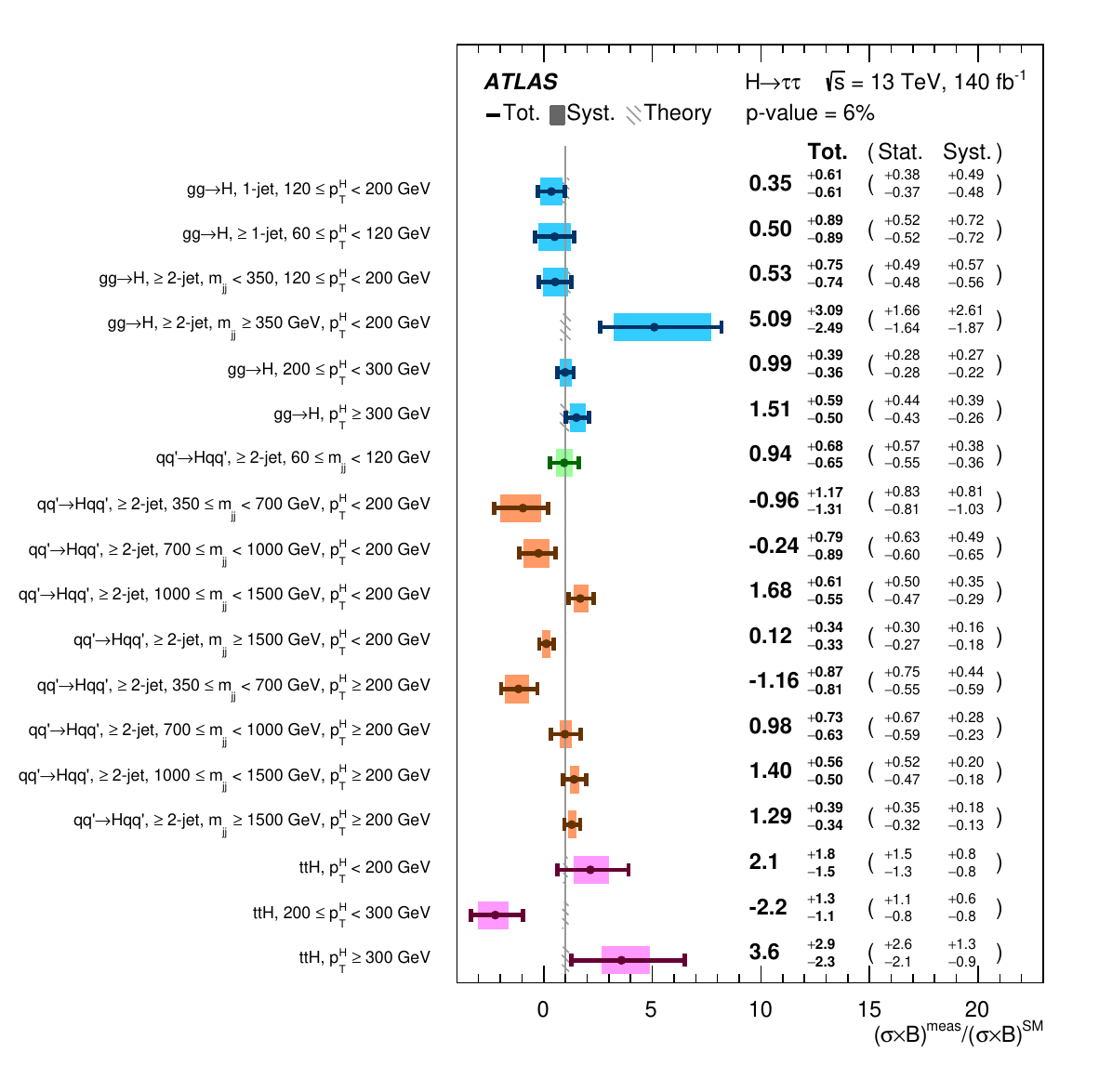}
\caption{
The measured values for $\sigma_{H}\times B$(\Htautau) relative to the SM
expectations in the simplified template cross-section measurement.
The error bars and shaded areas show the total uncertainty and systematic uncertainty in the measurements, respectively. }
\label{fig:stxs:18pois}
\end{center}
\end{figure}

\begin{figure}[htb!]
\begin{center}
\includegraphics[width=0.8\linewidth]{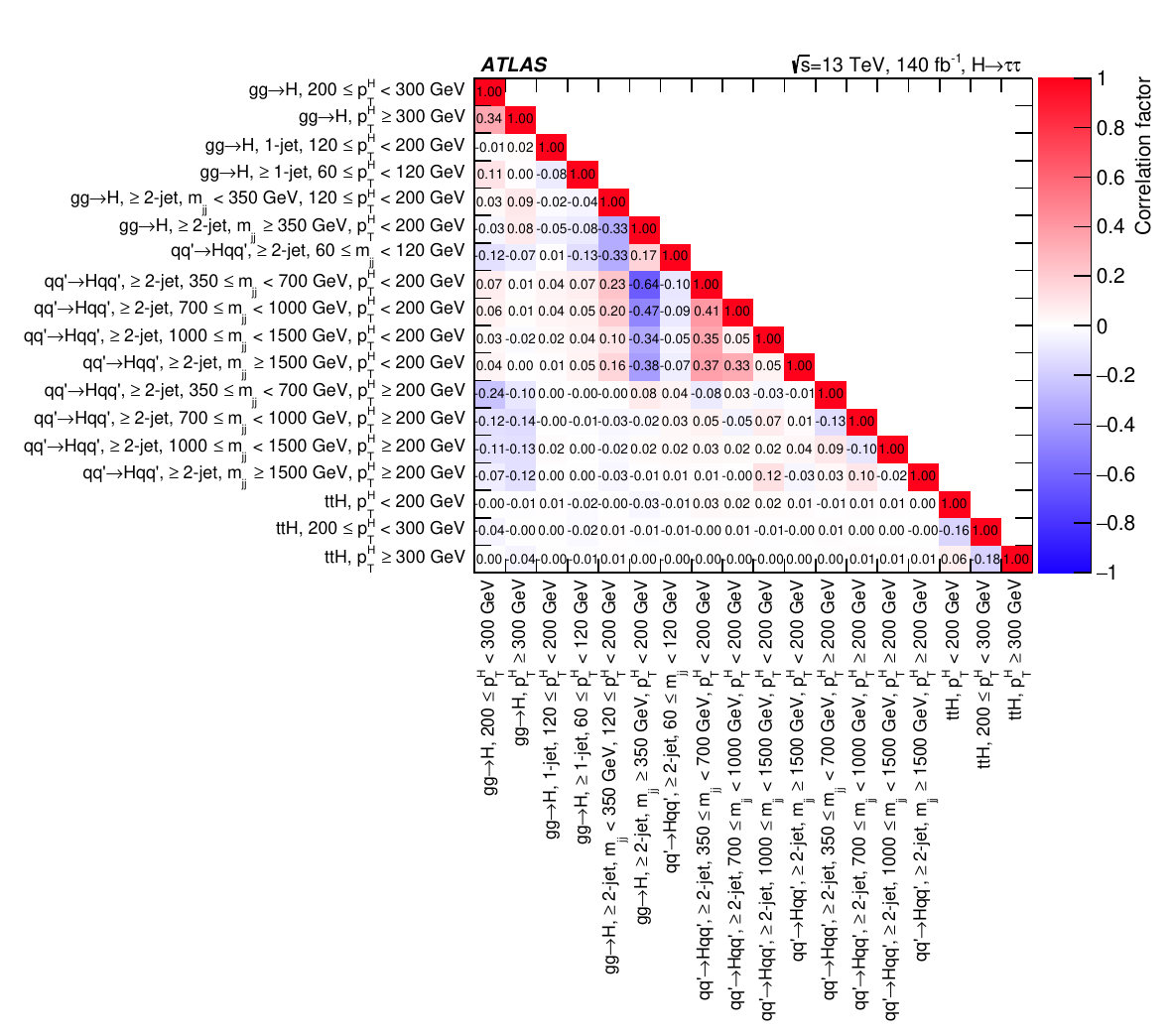}
\caption{
Measured linear correlation factors between each pair of parameters of interest in the simplified template cross-section measurement.
}
\label{fig:stxs:18pois_corr}
\end{center}
\end{figure}

\cref{fig:stxs:18pois} shows the results of the STXS measurement. The measurements in the inclusive ggF and VH regions are in good agreement with the SM and are similar to those in Ref. \cite{HIGG-2019-09} as the selections for these regions have not changed, although some of the theoretical uncertainties have been updated.
The eight bins used for VBF production show no significant deviation from the SM.
In the phase space with higher \pTH and/or higher \mjj the precision of the results is improved due to the smaller SM backgrounds in these regions, leading to a ratio of the measured cross-section to the SM expectation when $\pTH>200~\GeV$ ($\pTH<200~\GeV$) and $\mjj>1.5~\TeV$ of ${1.29}^{+0.39}_{-0.34}$ (${0.12}^{+0.34}_{-0.33}$).  This is the first measurement for the higher-\pTH~criteria, and the most precise for the lower-\pTH~criteria, in this corner of phase space.
As shown in \cref{fig:stxs:18pois_corr}, several VBF production POIs at low $\pTH$ are strongly anti-correlated with a single POI describing ggF production plus two jets, causing the negative values obtained for the VBF bins at $\pTH<200~\GeV$.
Overall, the $p$-value for compatibility of this result with the SM is 6\%. %

\begin{figure}[htb!]
\begin{center}
\includegraphics[width=0.5\linewidth]{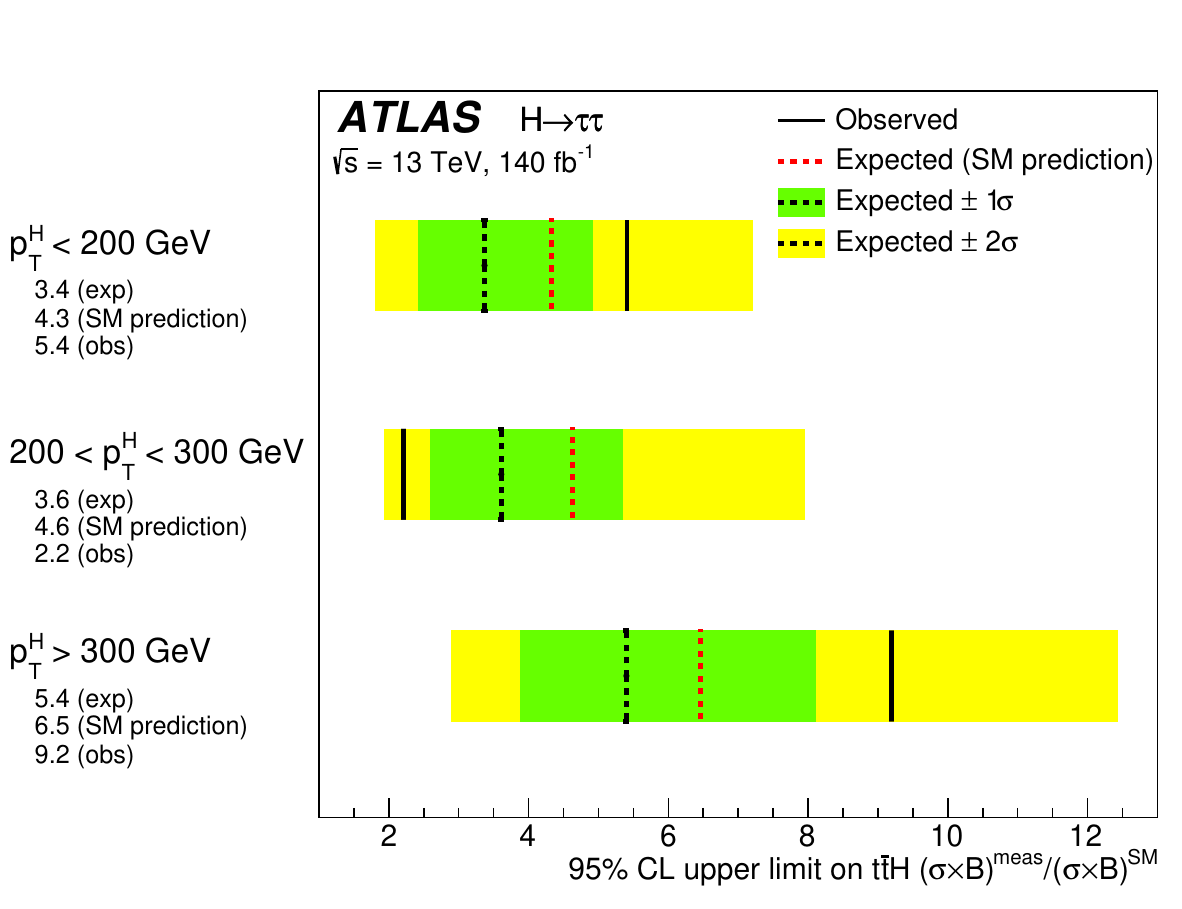}
\caption{
Upper limits at 95\% CL on simplified template cross-sections in the individual $\ttH$ $\pTH$ bins relative to their SM expectation in the \Htautau decay channel, derived using the CL$_\text{s}$ method.
The observed upper limits (solid black lines) are shown, together with the expected limits both in the background-only hypothesis (dotted black lines) and in the SM hypothesis (dotted red lines).
In the case of the expected limits in the background-only hypothesis, one- and two-standard-deviation uncertainty bands are also shown.
}
\label{fig:stxs:tthlimits}
\end{center}
\end{figure}

The results for $\ttH$ production are limited by statistical uncertainties, and they are used to derive cross-section upper limits at 95\% confidence level (CL) for $\ttH$ production in the STXS framework, using the CL$_\text{s}$ method~\cite{Junk:1999kv,Read:2002hq}, as shown in \cref{fig:stxs:tthlimits}.

\FloatBarrier

\section{Unfolded fiducial differential cross-section measurement results}
\label{sec:differential_results}

Another way to study the properties of VBF Higgs boson production is to measure its cross-section in a fiducial phase space that targets the VBF phase space defined by detector acceptance and selection requirements. The measured cross-sections are inclusive in the production process, thus minimizing SM assumptions. The measurements are corrected for detector effects and efficiencies (unfolded) using the combined SM MC prediction of VBF, ggF, $VH$ and $\ttH$ production.

The fiducial differential cross-section is defined in each particle-level kinematic bin, $i$, of the variable under study as shown in Eq.~(\ref{eq:diffxs}):

\begin{equation}
\sigma_i^{\textrm{fid}} = \frac{1}{\mathcal{L} \epsilon_i}\sum_j \mathcal{M}^{-1}_{ij} f_j (N^{\mathrm{data}}_{j} - N^{\mathrm{bkg}}_{j}) \Hquad ,
\label{eq:diffxs}
\end{equation}

where $j$ is the reconstruction-level bin index, $\mathcal{L}$ is the integrated luminosity, and $\mathcal{M}$ is the migration matrix for events passing both reconstruction- and fiducial-level selections. $\mathcal{M}$ gives the probability that an event in particle-level bin $i$ is reconstructed in reconstruction-level bin $j$. The selection efficiency $\epsilon_i$ is defined as the number of events passing both the fiducial- and reconstruction-level selections divided by the number of events passing the fiducial-level selection. The out-of-acceptance correction $f_j$ is defined as the number of events passing both the fiducial- and reconstruction-level selections divided by the number of events that pass the reconstruction-level selection. This is multiplied by the number of signal events in the SRs, defined as the number of events observed in data ($N^{\mathrm{data}}_{j}$) after the estimated total number of background events ($N^{\mathrm{bkg}}_{j}$) is subtracted. $\mathcal{M}$, $\epsilon_i$, and $f_j$ are determined using the SM signal MC prediction.

In the following, the fiducial phase space is defined, the measured variables are introduced, the likelihood fit model is discussed, and the results of the fiducial differential cross-section measurement are presented.

\subsection{Fiducial phase space}

The fiducial differential cross-sections are evaluated in a fiducial phase space. This phase space is defined by particle-level kinematic criteria which closely follow the differential inclusive VBF selection applied at reconstruction level and are defined in Tables~\ref{tab:selection:baseline} and \ref{tab:selection:signalregions}. The fiducial phase space, in which VBF production constitutes approximately 83\% of the Higgs boson signal, and ggF production approximately 17\%, is defined in Table~\ref{tab:fiducial:perchannel}.  The VBF, ggF, $VH$, and $\ttH$ Higgs boson production modes are combined as signal, and their relative contributions are fixed to the SM predictions.

For the object definition, only stable final-state particles (with a mean lifetime $c\tau > 10\,\mathrm{mm}$) are considered at particle level.
Electrons and muons are required to not originate from hadrons or hadronic decay products. Their four-momenta are corrected by adding the momenta of final-state radiation photons emitted in a cone of size $\Delta R = 0.1$ around the lepton direction. Hadronically decaying $\tau$-leptons (denoted by \tautruth at particle level) are reconstructed by identifying all stable $\tau$-lepton decay particles except for neutrinos and summing their four-momenta. All leptons are required to have $|\eta| < 2.5$, and electrons are excluded from the LAr transition region $1.37 < |\eta| < 1.52$. Jets are defined by clustering all stable particles using the anti-\kt algorithm with a radius parameter of $R = 0.4$. The clustering algorithm excludes prompt leptons and Higgs boson decay products. Jets are required to be separated from \tautruth ($\Delta R > 0.4$) and light leptons ($\Delta R > 0.2$), and are required to have $|\eta| < 4.5$. The missing transverse momentum is defined as the vector sum of the transverse momenta of all neutrinos that do not originate from the decay of a hadron.

%
%
%
%
%
%
%
%
%
%
%
%
%
%
%
%
%
%
%
%
%
%
%
%
%
%
%
%
%
%

\begin{table}[tbhp]
\begin{center}
\caption{
Summary of the fiducial particle-level event selection for all channels. The requirements on light-lepton \pT and \tautruth \pT follow those reported in \cref{tab:selection:baseline}.
}
\label{tab:fiducial:perchannel}
\scalebox{0.72}{
\begin{tabular}{r|c|c|c}
\toprule
& \temu & \tlhad  & \thadhad \\
\midrule
&
&
&  \\
\textbf{Per Channel} &
&
&  \\
Object counting       &  $N_{e}$ = 1, $N_{\mu} = 1$,  $N_{\tautruth} = 0$
&  $N_{e/\mu} = 1$, $N_{\tautruth} = 1$
&  $N_{e/\mu} = 0$, $N_{\tautruth} = 2$\\
&
&
& \\
\pT cut               & $e$/$\mu$ : \pT cut \SIrange{10}{27.3}{GeV}
& $e$/$\mu$ : \pT cut \SIrange{27.0}{27.3}{GeV},
& \\
&
& \tautruth: $\pT > \SI{30}{\GeV}$
& \tautruth: $\pT > \SI{40}, \SI{30}{\GeV}$ \\
&
&
& \\
Kinematics            & $m_{\tau\tau}^{\mathrm{coll}} > m_Z - \SI{25}{\GeV}$
& $\mT < \SI{70}{\GeV}$
& \\
& $30 < m_{e\mu} < \SI{100}{\GeV}$
&
& \\
Angular          & $\Delta R_{e\mu} < 2.0$, $|\Delta\eta_{e\mu}| < 1.5$
& $\Delta R_{\ell\tautruth} < 2.5$, $|\Delta\eta_{\ell\tautruth}| < 1.5$
& $0.6 < \Delta R_{\tautruth\tautruth} < 2.5$ \\
&
&
& $|\Delta\eta_{\tautruth\tautruth}| < 1.5$\\
&
&
& \\
$x_{1}$ and $x_{2}$         & $0.1 < x_{1} < 1.0$, $0.1 < x_{2} < 1.0$
& $0.1 < x_{1} < 1.4$, $0.1 < x_{2} < 1.2$
& $0.1 < x_{1} < 1.4$, $0.1 < x_{2} < 1.4$ \\
&
&
&  \\
\midrule
\multicolumn{4}{c}{ } \\
\midrule
& \multicolumn{3}{c}{ } \\
\textbf{Common selection} & \multicolumn{3}{c}{leading jet $\pT > \SI{40}{\GeV}$, sub-leading jet $\pT > \SI{30}{\GeV}$} \\
& \multicolumn{3}{c}{$\MET > \SI{20}{\GeV}$} \\
& \multicolumn{3}{c}{Opposite charge of $\tau$-decay products}  \\
& \multicolumn{3}{c}{$\mjj > \SI{600}{\GeV}$, $|\detajj| > 3.4$, $\ptjj > \SI{30}{\GeV}$} \\
& \multicolumn{3}{c}{$\eta(j_{0}) \times \eta(j_{1}) < 0$} \\
& \multicolumn{3}{c}{lepton centrality: visible decay products of the $\tau$-leptons between VBF jets} \\
& \multicolumn{3}{c}{$\pTHjj  < \SI{50}{\GeV}$} \\
& \multicolumn{3}{c}{ } \\

\bottomrule
\end{tabular}
}
\end{center}
\end{table}


\subsection{Differential variables}
The differential cross-sections are measured in the fiducial phase space as a function of variables sensitive to the kinematics of VBF Higgs boson production, with the goal of validating the SM as well as providing sensitivity to physics beyond the SM. The variables related to jets are the transverse momentum of the leading jet, $\pTj$, and the signed angle (sorted by rapidity of the jets) in the transverse plane of the two jets, $\dPhiJJsign$, which also provides a way to test the charge (C) and parity (P) of the Higgs boson. The kinematics of the Higgs boson is tested by measuring its transverse momentum ($\pTH$). As in the STXS measurement, the NN reconstruction of $\pTH$, introduced in Section~\ref{subsec:vbfcategorisation}, is used at reconstruction level.
To increase the sensitivity to the CP property of the Higgs boson, $\dPhiJJsign$ is also measured as a function of $\pTH$.
The binning of each differential distribution is driven by the size of the available data sample and is reported in Table~\ref{tab:sel_fid:binning}.

\begin{table}[h]

\centering
\caption{Binning of the differential variables.}
\begin{tabular}{r|c|c|c|c}
\toprule
&	  $\pTj$ [GeV]  &     $\pTH$ [GeV]   & $\dPhiJJsign$         & $\dPhiJJsign$ vs $\pTH$ [GeV] \\ \midrule
Bin 1 &     $[40,95]$      &   ~~~~$[0,110]$   & $[-\pi,-\pi/2]$   & $\dPhiJJsign <0$ $\&$ $\pTH < 200 $   \\
Bin 2 &     ~~$[95,130]$    &   $[110,150]$ & $[-\pi/2,0] $     & $\dPhiJJsign >0$ $\&$ $\pTH < 200 $ \\
Bin 3 &     $[130,180]$    &   $[ 150,200]$ & $[0,\pi/2]$      & $\dPhiJJsign <0$ $\&$ $\pTH > 200 $ \\
Bin 4 &      $[180,500]$   &   $[200,550]$ & $[\pi/2,\pi]$     & $\dPhiJJsign >0$ $\&$ $\pTH > 200 $ \\ \bottomrule
\end{tabular}
\label{tab:sel_fid:binning}
\end{table}

The migration matrix is highly diagonal for all variables considered, with values above 75\%, 85\%, 95\% and 90\% for the \pTj, \pTH, \dPhiJJsign, and \dPhiJJsign vs \pTH distributions. This is due to the binning being coarse compared to experimental resolution effects. The selection efficiency, $\epsilon_i$, includes the effect of object reconstruction efficiencies and event-level selections. These are 25\%, 15\%, and 43\% for the $\tau_{\textrm{lep}}\tau_{\textrm{had}}$, $\tau_{\textrm{had}}\tau_{\textrm{had}}$, and $\tau_e\tau_\mu$ channels respectively. The out-of-acceptance corrections, $f_j$, range from 74\% to 83\% across the channels.

%
%
%
%
%
%
%
%
%
%
%
%
%
%
%
%
%
%
%
%
%
%
%
%
%
%
%
%
%


%

\subsection{Differential cross-section measurement fit model}
\label{sec:differential_results:fitModel}
For the differential cross-section measurement, the likelihood function is built with the following SRs and CRs:

\begin{itemize}
\item Eight SRs for the VBF category in each channel, based on the BDT-based splitting, as described in \cref{subsec:vbfcategorisation}, plus the different bins reported in \cref{tab:sel_fid:binning}.
\item Two \Zttjets control regions in each channel, one for each inclusive VBF\_0 and VBF\_1  SR, defined using the kinematic embedding.
\item A CR enriched in the \ttbar process, defined for each inclusive category in the \temu and \tlhad channels.
\end{itemize}

\begin{figure}[htb!]
\begin{center}
\begin{tabular}{c}
\includegraphics[width=0.9\linewidth]{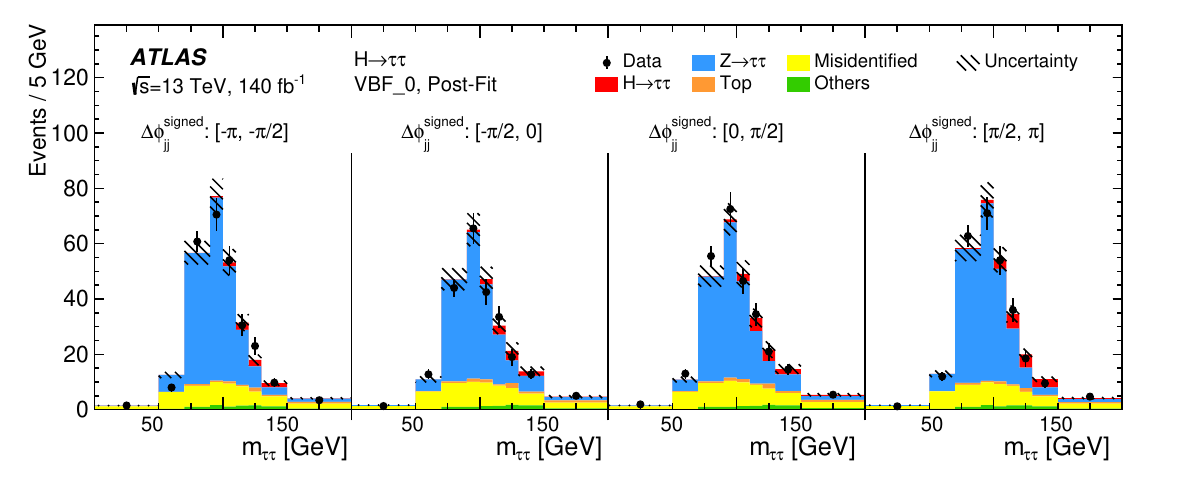} \\
(a) \\
\includegraphics[width=0.9\linewidth]{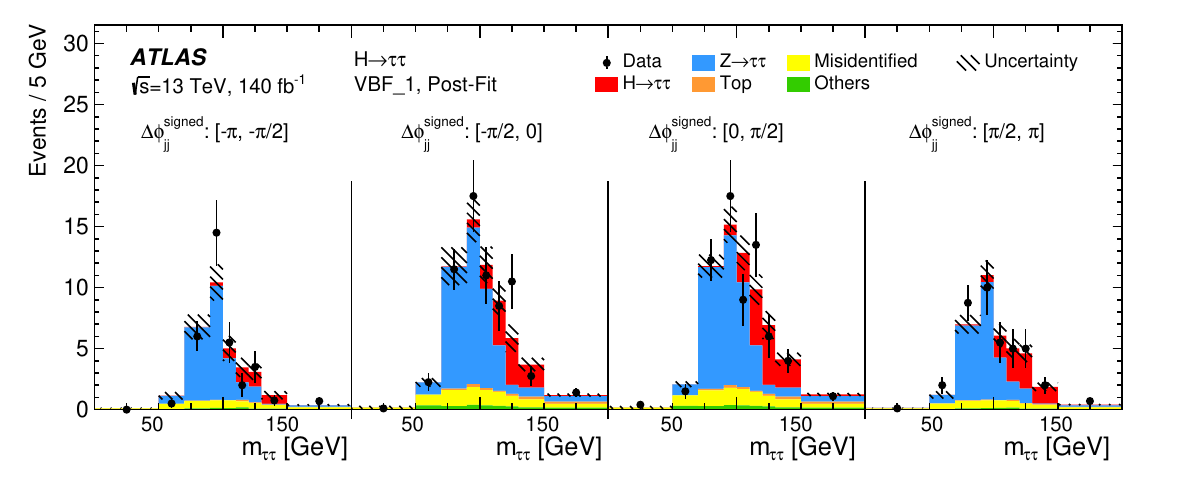} \\
(b) \\
\end{tabular}
\caption{
Distribution of the reconstructed $\tau\tau$ invariant mass ($m_{\tau\tau}$) for
all events in the (a)  VBF\_0 and (b)  VBF\_1 signal
regions,  shown for each bin of the $\dPhiJJsign$ observable. The observed
Higgs boson signal is scaled by the cross-section measured in each bin.
Entries with values outside the $x$-axis
range are shown in the last bin of each distribution.
The prediction for each sample is determined from the likelihood fit performed to
measure the fiducial differential cross-section.
\label{fig:diff:dphijj}}
\end{center}
\end{figure}

Similarly to the STXS cross-section measurement set-up, the $m_{\tau\tau}$ distribution is used to obtain information about the cross-section relative to the SM prediction, and a single-bin histogram is used for each CR to constrain the background normalization.
The detector-level $m_{\tau\tau}$ distributions for each bin of the $\dPhiJJsign$ observable (specified in \cref{tab:sel_fid:binning}) are shown in \cref{fig:diff:dphijj} and compared with the MC templates after the likelihood fit. In these distributions, all three analysis channels are combined in the VBF\_0 and VBF\_1 SRs.

The differential measurements are performed by using the profile-likelihood unfolding~\cite{TOPQ-2020-20} in the signal regions. The free parameters of the fit are the Higgs boson production cross-sections %
in each bin of the distribution at particle level. No regularization is applied, due to the small bin-to-bin migrations for all variables considered. The performance of the unfolding procedure is tested for possible biases from the choice of MC model. It is verified that it reproduces altered shapes well within the statistical precision of the measurements.

\subsection{Differential cross-section measurement results}

\begin{figure}[htb!]
\begin{center}
\begin{tabular}{cc}
\includegraphics[width=0.49\linewidth]{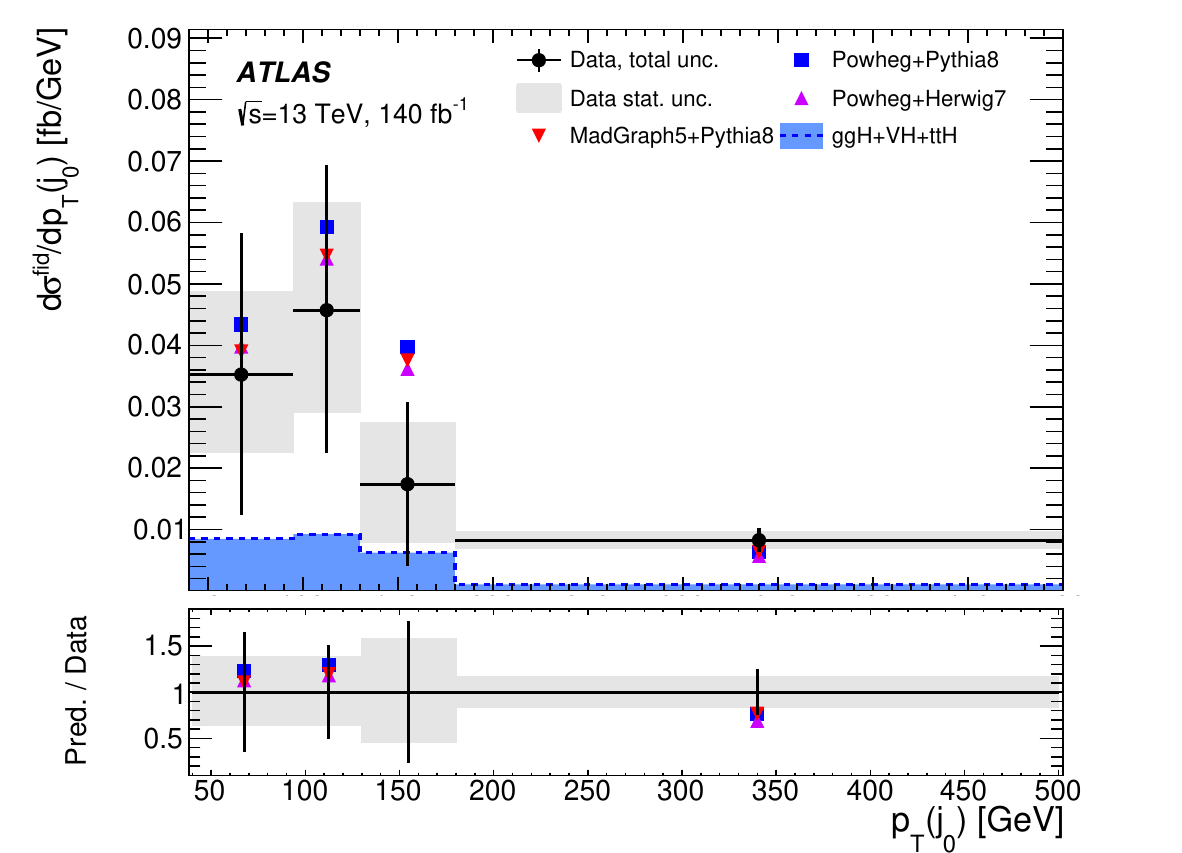} &
\includegraphics[width=0.49\linewidth]{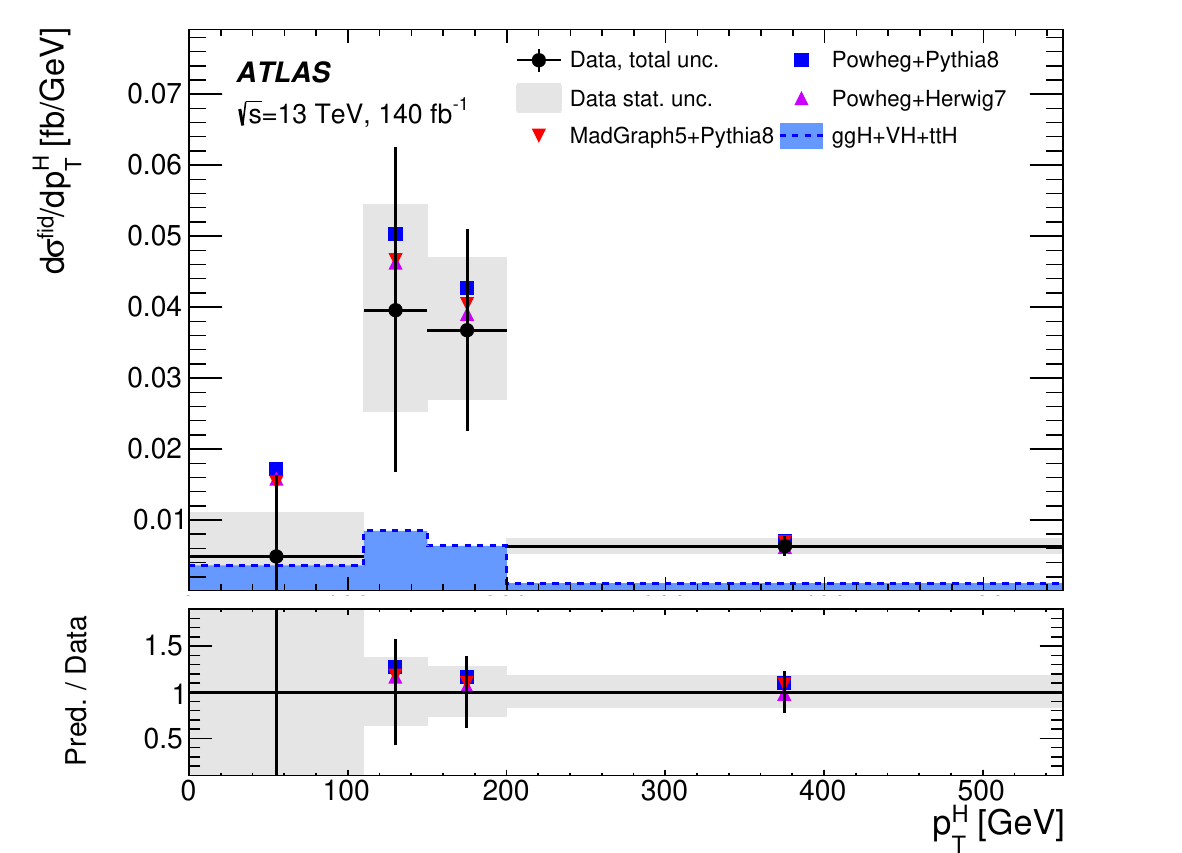} \\
(a) & (b) \\
\includegraphics[width=0.49\linewidth]{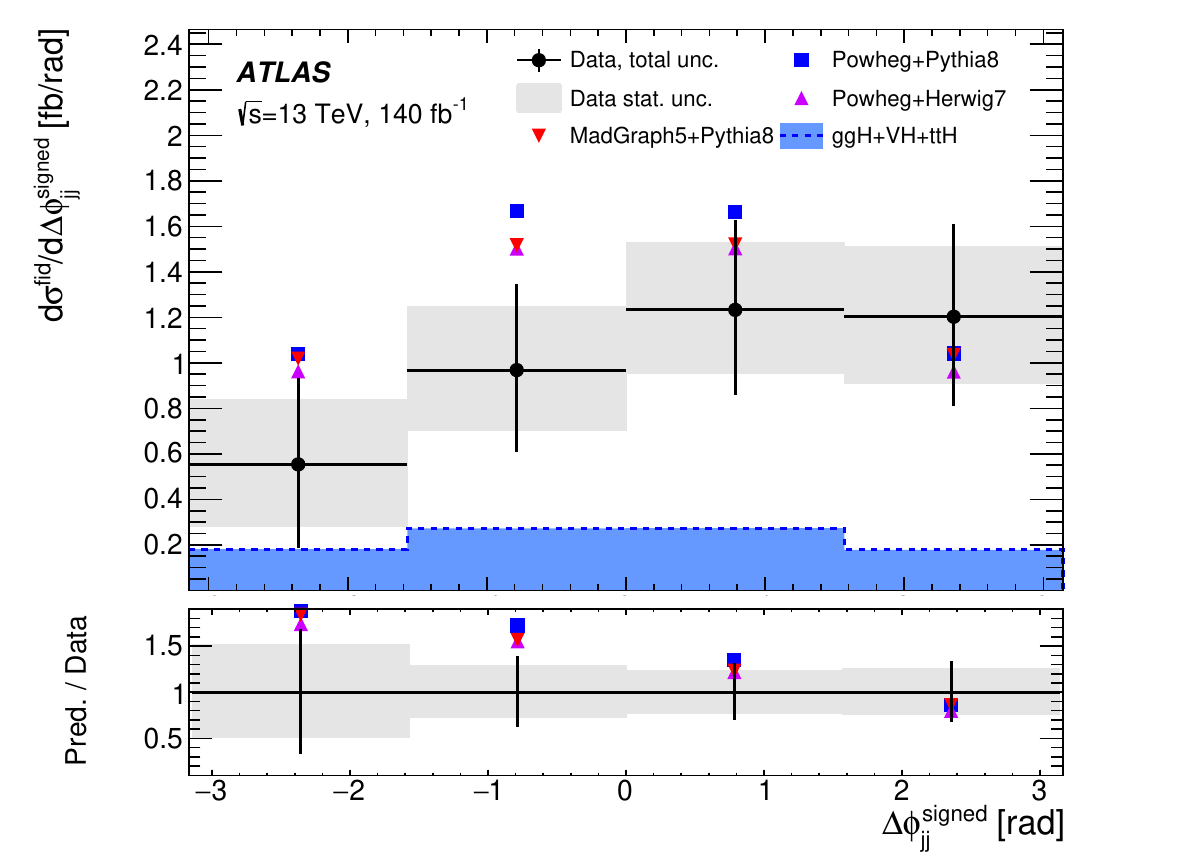} &
\includegraphics[width=0.49\linewidth]{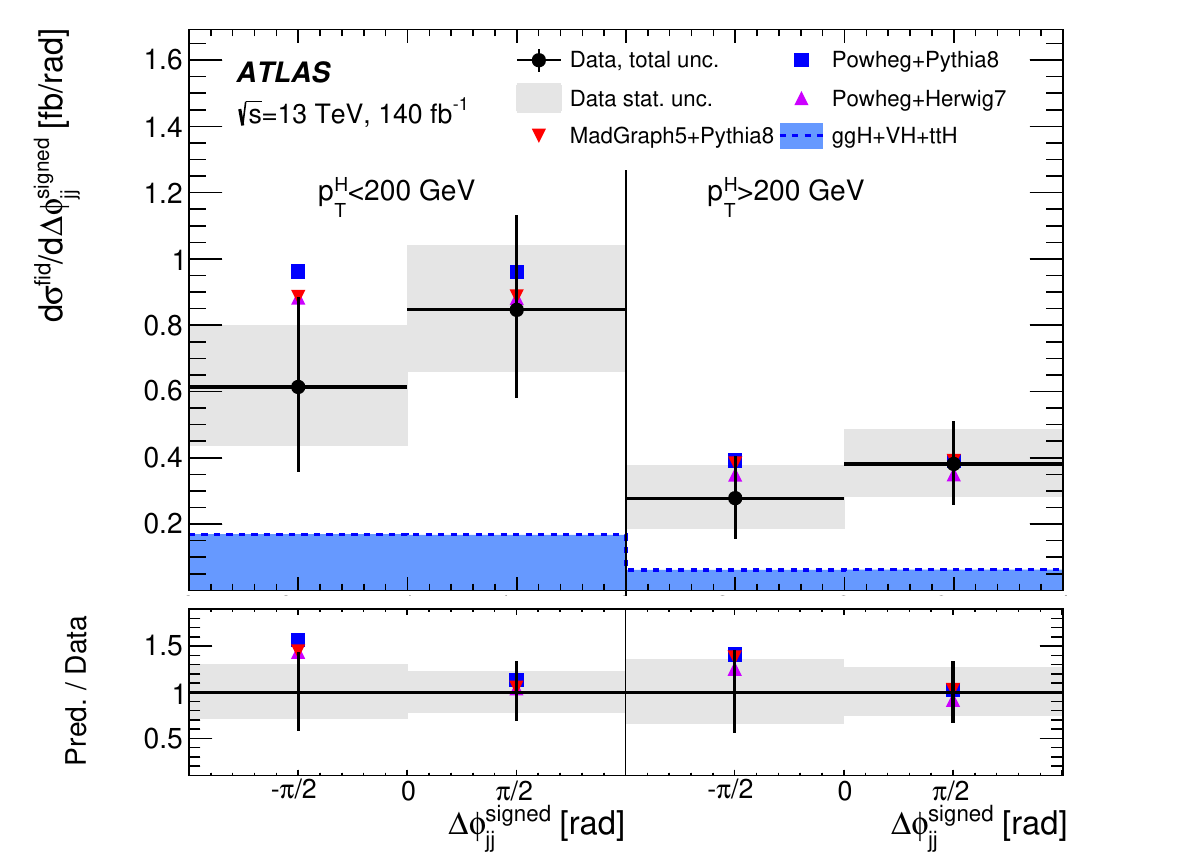} \\
(c) & (d) \\
\end{tabular}
\caption{
Measured fiducial differential cross-sections for (a) $\pTj$, (b) $\pTH$, (c) $\dPhiJJsign$, and (d) $\dPhiJJsign$ vs $\pTH$.  The measurements are compared with particle-level SM predictions from the \POWPY[8], \POWHER[7] and \MGNLOPY[8] generators for the combined VBF, ggF, $VH$, and $\ttH$ Higgs boson production modes. The shaded box around each data point shows the statistical uncertainty, while the total uncertainty is indicated by the error bar. The contribution from the ggF, $VH$, and $\ttH$ production modes as predicted by \POWPY[8] is also shown.
The bottom panel shows the ratio of different predictions to the data, with the error bars and shaded bands representing the total and statistical uncertainties of the measurements, respectively.
\label{fig:diff:unfolded}}
\end{center}
\end{figure}

The fiducial differential cross-sections as a function of the variables of interest are shown in \cref{fig:diff:unfolded}.
The per-bin precision of the differential cross-section measurements is typically
25\%--50\%, except for the lowest $\pTH$ bin, where it reaches 300\%.
The experimental precision is dominated by the statistical uncertainties in most of the bins,  particularly at high values of $\pTH$ and $\pTj$. Leading systematic uncertainties, depending on the measured bin, are the limited statistics of the background model, experimental uncertainties for jets and $\tau$-leptons, and uncertainties in the fake-lepton background estimation.

The data are compared with SM predictions from the \POWPY[8], \POWHER[7] and \MGNLOPY[8] generators for the VBF, ggF, $VH$, and $\ttH$ Higgs boson production modes, which are combined according to their SM branching ratios. The theoretical uncertainty in these predictions is much smaller than the experimental uncertainty in the measurements.
The \textsc{Rivet} analysis library~\cite{rivet} is used to apply the fiducial selections and calculate the observables to obtain the differential cross-section predictions.
Overall, there is good agreement between measured cross-sections and the SM predictions from all generators, with only a few bins showing discrepancies of more than one standard deviation.

%
%
%
%
%
%
%
%
%
%


\clearpage
\subsection{Interpretation of the differential cross-section measurement in an Effective Field Theory}
The unfolded distributions are sensitive to effects beyond-the-SM (BSM) contributions in VBF Higgs boson production. This motivates the interpretation of the results in a generic \textit{Effective Field Theory} (EFT) that parameterizes these potential BSM effects. Anomalous higher-dimensional operators are added to the SM Lagrangian to encapsulate the effects of BSM physics at higher scales, as shown in Eq.~(\ref{eqn:eft}):
\begin{equation}
\label{eqn:eft}
\mathcal{L}_{\textrm{EFT}} = \mathcal{L}_{\textrm{SM}} + \sum_d\sum_i \frac{c_i^{(d)}}{\Lambda^{d-4}}\mathcal{O}_i^{(d)}, \textrm{ for }d>4 \Hquad .
\end{equation}
The parameters $c_i^{(d)}$, referred to as Wilson coefficients, specify the strength of the anomalous interactions introduced by their corresponding operators, $\mathcal{O}_i^{(d)}$, where the index $i$ runs over all possible operators, and $\Lambda$ sets the scale of the BSM physics. Only dimension-six operators are considered since the dimension-five and dimension-seven operators violate lepton- or baryon-number conservation, and the impact of higher-dimensional operators is expected to be further suppressed~\cite{Hays:2018zze}. In this scenario, only the ratio $c_i^{(d=6)}/\Lambda^2$ determines the strength of the BSM physics and $\Lambda$ is chosen to be 1~\TeV by convention.

In the Warsaw basis~\cite{Grzadkowski:2010es} of the \textit{Standard Model Effective Field Theory} (SMEFT) formalism~\cite{Brivio:2017btx}, there are three CP-even and three CP-odd operators that directly affect the interactions between the Higgs boson and the vector bosons; they are listed alongside their corresponding Wilson coefficients in Table \ref{tab:diffEFT:Wilson}. The change in the cross-section for each kinematic bin as a function of the Wilson coefficients follows a second-order polynomial as there is an interference term which depends linearly on the Wilson coefficient and a quadratic term that is purely from BSM physics as shown in Eq.~(\ref{eq:diffEFT:xsec}):
\begin{equation}
\sigma_{\textrm{SM+EFT}} \propto |\mathcal{M}_{\textrm{SM}} + \sum_i \frac{c_i}{\Lambda^2}\mathcal{M}_i |^2 = |\mathcal{M}_\textrm{SM}|^2 + 2\sum_i \frac{c_i}{\Lambda^2} \mathcal{R}(\mathcal{M}_\textrm{SM}^{*}\mathcal{M}_i) + \sum_{i,j} \frac{c_i c_j}{\Lambda^4}\mathcal{R}(\mathcal{M}_i^{*}\mathcal{M}_j) \Hquad .
\label{eq:diffEFT:xsec}
\end{equation}
Therefore, for each Wilson coefficient, the dependence in each kinematic bin, $k$, can be parameterized using the form
\begin{equation}
\sigma_{\textrm{SM+EFT}}^k = \sigma_{\textrm{SM}}^k\left(1 + \alpha_{ik} \frac{c_i}{\Lambda^2} + \beta_{ik} \left(\frac{c_i}{\Lambda^2}\right)^2\right) \Hquad .
\label{eq:diffEFT:alphaBeta}
\end{equation}
The coefficients $\alpha_{ik}$ and $\beta_{ik}$ are determined from simulated samples with different values of $c_i$ produced using \MGNLO[2.9.5] with the SMEFTsim~\cite{Brivio:2017btx,Brivio:2020onw} model in the $U(3)^5$ flavour-symmetric limit with $\Lambda = 1$~\TeV. The events were generated at LO using the \NNPDF[3.0nlo] PDF set and were showered using \PYTHIA[8.245] with the A14 tune. The dependence was determined relative to a SM sample produced using the same set-up and validated against the sample described in  \cref{sec:samples}. These fractional changes determined from these BSM samples are then applied to the VBF cross-section predictions from the \POWHEGBOX[v2] sample showered with \PYTHIA[8.230] also described in  \cref{sec:samples}. For the small contributions from Higgs boson production processes other than VBF, the various other samples described in  \cref{sec:samples} are used.

\begin{table}
\centering
\caption{The three CP-even and three CP-odd operators and their corresponding Wilson coefficients which directly impact the interactions between the Higgs boson and vector bosons.}
\label{tab:diffEFT:Wilson}
\begin{tabular}{|c|c|c|c|}
\hline
&\multicolumn{3}{c|}{CP-even}\\ \hline
Operator $\mathcal{O}_i^{(d=6)}$ & $H^\dagger H W^n_{\mu\nu} W^{n\mu\nu}$ & $H^\dagger H B_{\mu\nu} B^{\mu\nu}$ & $H^\dagger\tau^n H W^n_{\mu\nu} B^{\mu\nu}$ \\ \hline
Wilson coefficient & $c_{HW}$ & $c_{HB}$ & $c_{HWB}$\\ \hline
&\multicolumn{3}{c|}{CP-odd}\\ \hline
Operator $\mathcal{O}_i^{(d=6)}$ & $H^\dagger H \tilde{W}^n_{\mu\nu} W^{n\mu\nu}$ & $H^\dagger H \tilde{B}_{\mu\nu} B^{\mu\nu}$ & $H^\dagger\tau^n H \tilde{W}^n_{\mu\nu} B^{\mu\nu}$ \\ \hline
Wilson coefficient & $c_{H\tilde{W}}$ & $c_{H\tilde{B}}$ & $c_{H\tilde{W}B}$\\ \hline
\end{tabular}
\end{table}

The linear term that is determined by $\alpha$~comes from the interference of the new physics with the SM matrix element and enters at the lowest order in $\mathcal{O}(\Lambda^{-2})$. The second, quadratic, term is purely from the new-physics contribution and enters at $\mathcal{O}(\Lambda^{-4})$. However, potential terms from dimension-8 operators interfering with the SM would also enter at the same order in $1/\Lambda$. Therefore, for the validity of the EFT approach only considering dimension-6 operators, it is desirable that the contribution from the quadratic term is small and the results are dominated by the contributions from the linear term which enters at $\mathcal{O}(\Lambda^{-2})$. To assess this, when considering each of the operators in Table~\ref{tab:diffEFT:Wilson}, measurements are made considering the full form of Eq.~(\ref{eq:diffEFT:alphaBeta}), and also excluding the quadratic term.

\begin{figure}[htb!]
\begin{center}
\begin{tabular}{cc}
\includegraphics[width=0.49\linewidth]{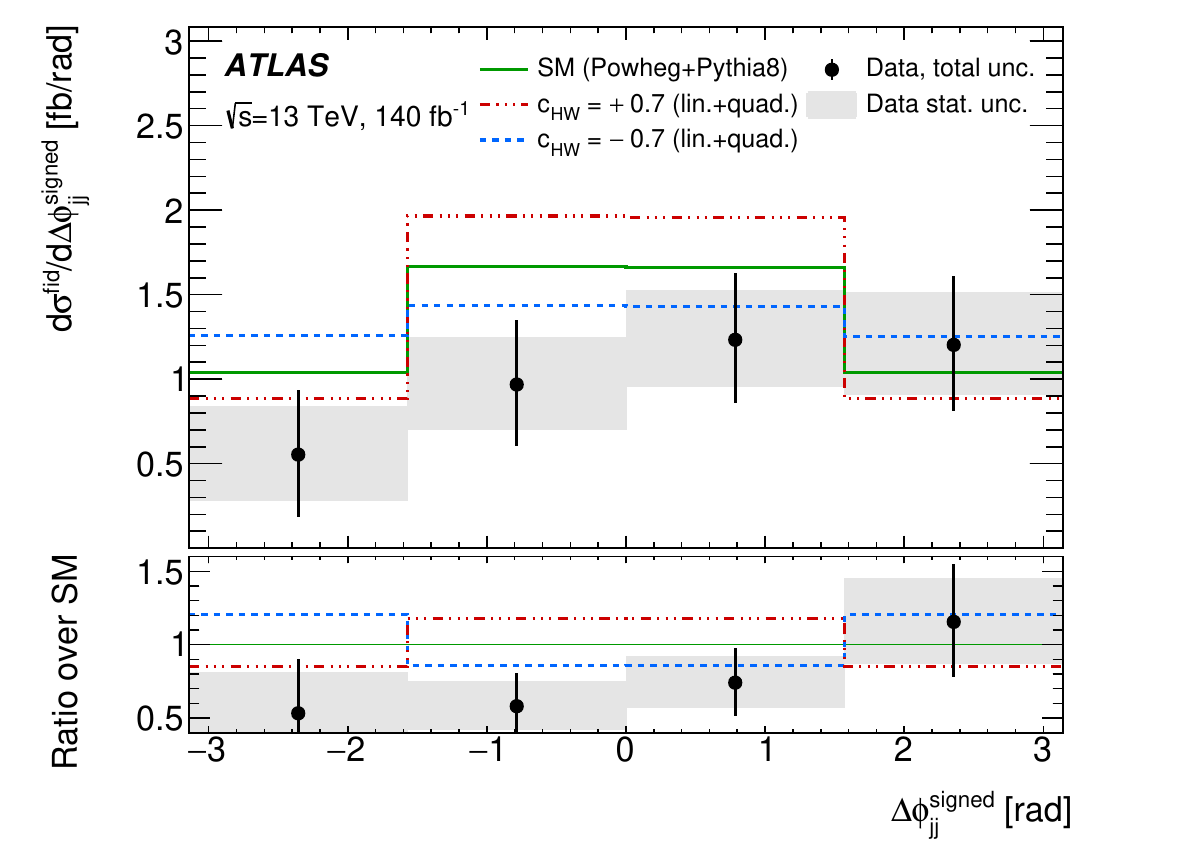} &
\includegraphics[width=0.49\linewidth]{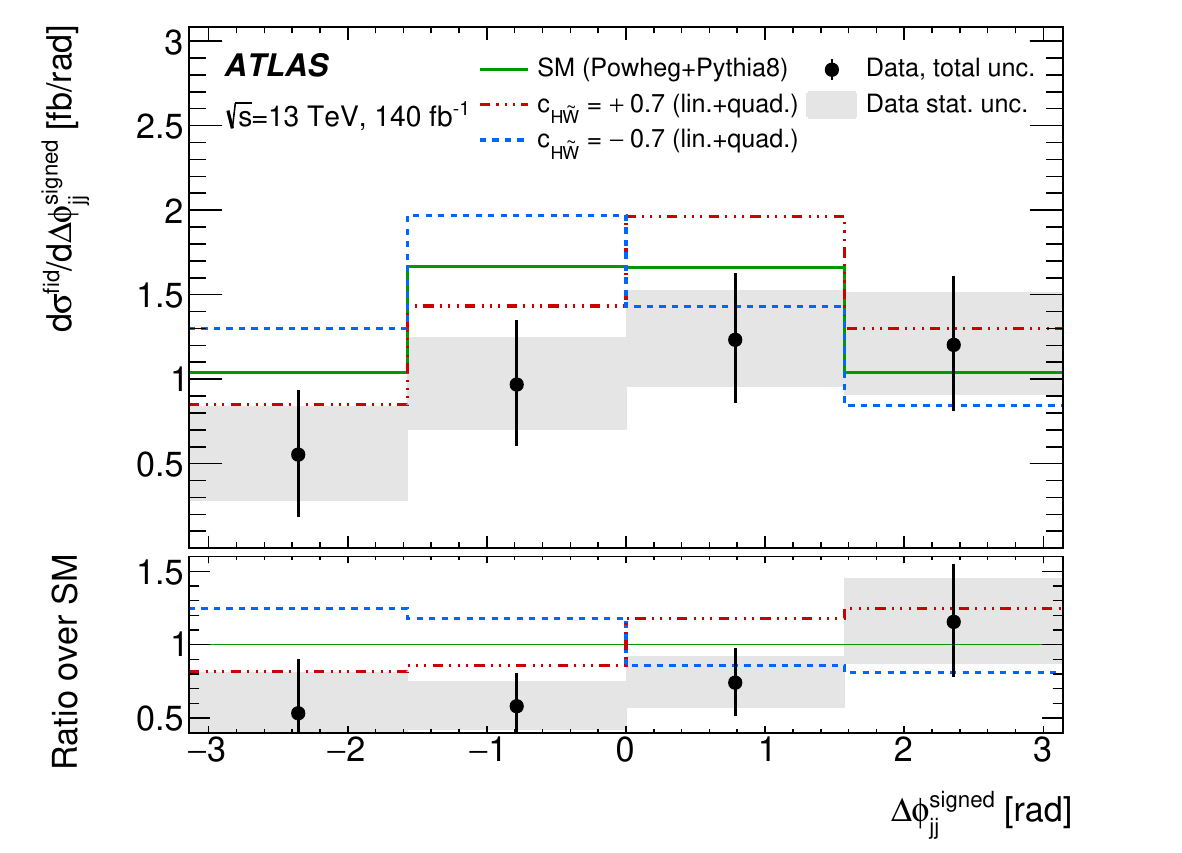} \\
(a) & (b) \\
\includegraphics[width=0.49\linewidth]{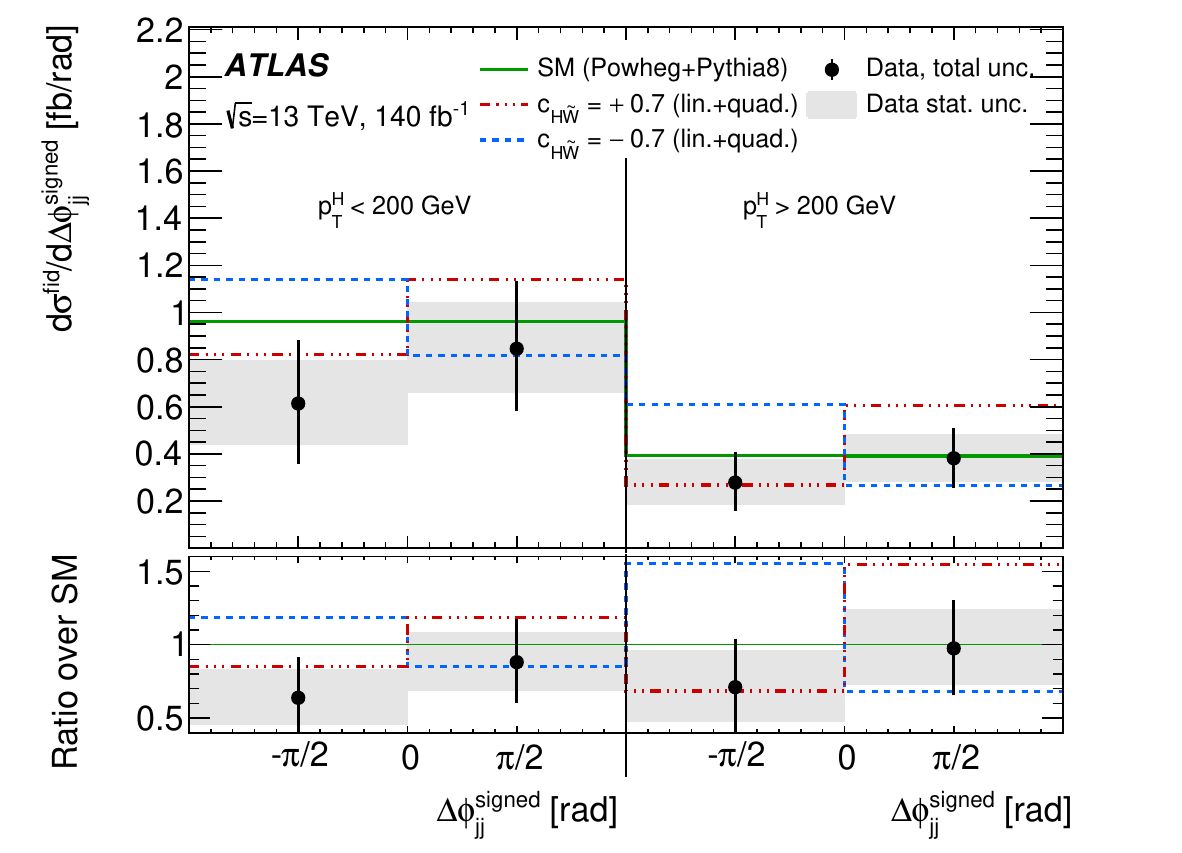} &
\includegraphics[width=0.49\linewidth]{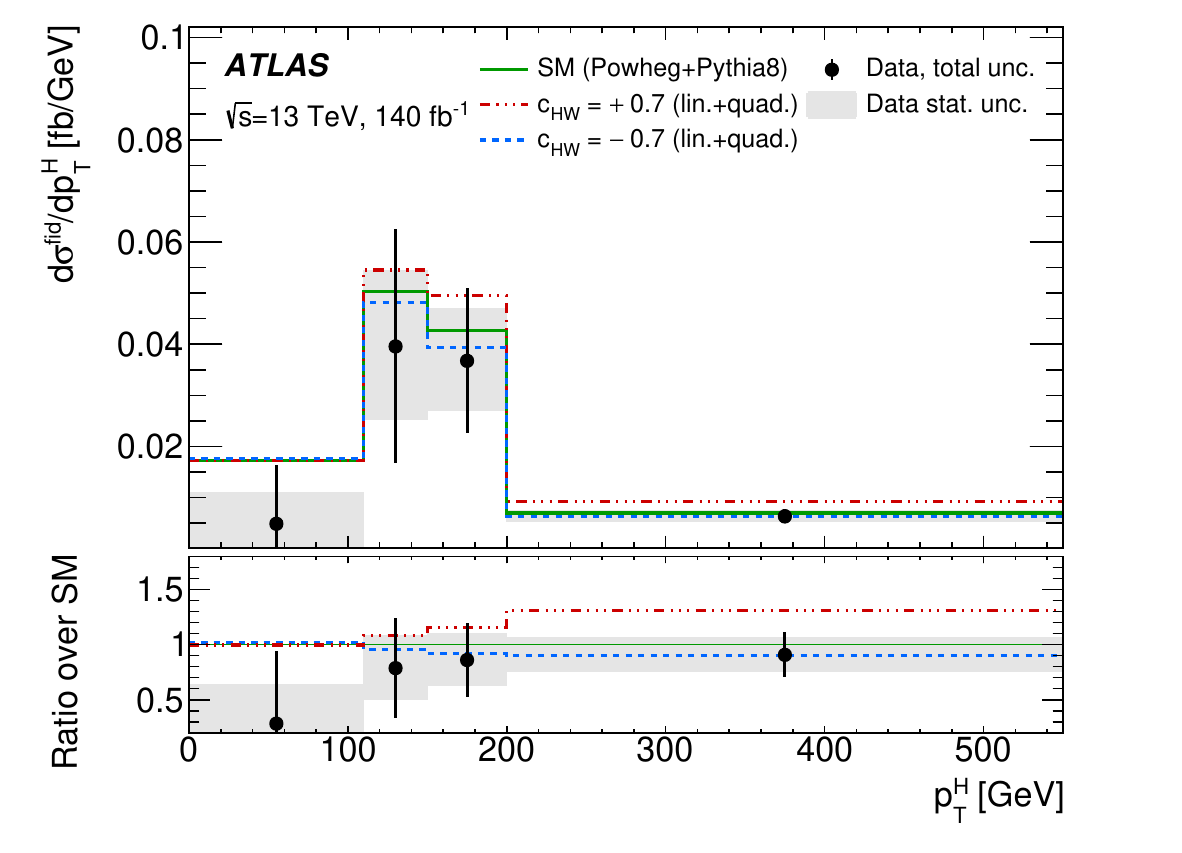} \\
(c) & (d) \\
\end{tabular}
\caption{
The unfolded data compared with the nominal SM predictions from \POWPY[8], as well as the predictions for several values of Wilson coefficients in the SMEFT, for (a,b) \dPhiJJsign, (c) \dPhiJJsign vs \pTH and (d) \pTH.
\label{fig:diffEFT:impact}}
\end{center}
\end{figure}

Figure~\ref{fig:diffEFT:impact} shows the impact of several values of $c_{HW}$ and $c_{H\tilde{W}}$ on the distributions of \dPhiJJsign, \pTH, and \dPhiJJsign vs \pTH, demonstrating that these variables are sensitive to BSM effects. Due to the CP-odd nature of the operator associated with $c_{H\tilde{W}}$, an asymmetric impact is seen in \dPhiJJsign and additionally it is seen that the relative magnitude of the effect is enhanced by the cut on \pTH in the two-dimensional distribution. For $c_{HW}$, effects are seen in both the \dPhiJJsign and \pTH distributions and, %
contrary to $c_{H\tilde{W}}$, the effects are symmetric in \dPhiJJsign due to the CP-even nature of the associated operator.

\begin{figure}[htb!]
\begin{center}
\begin{tabular}{cc}
\includegraphics[width=0.4\linewidth]{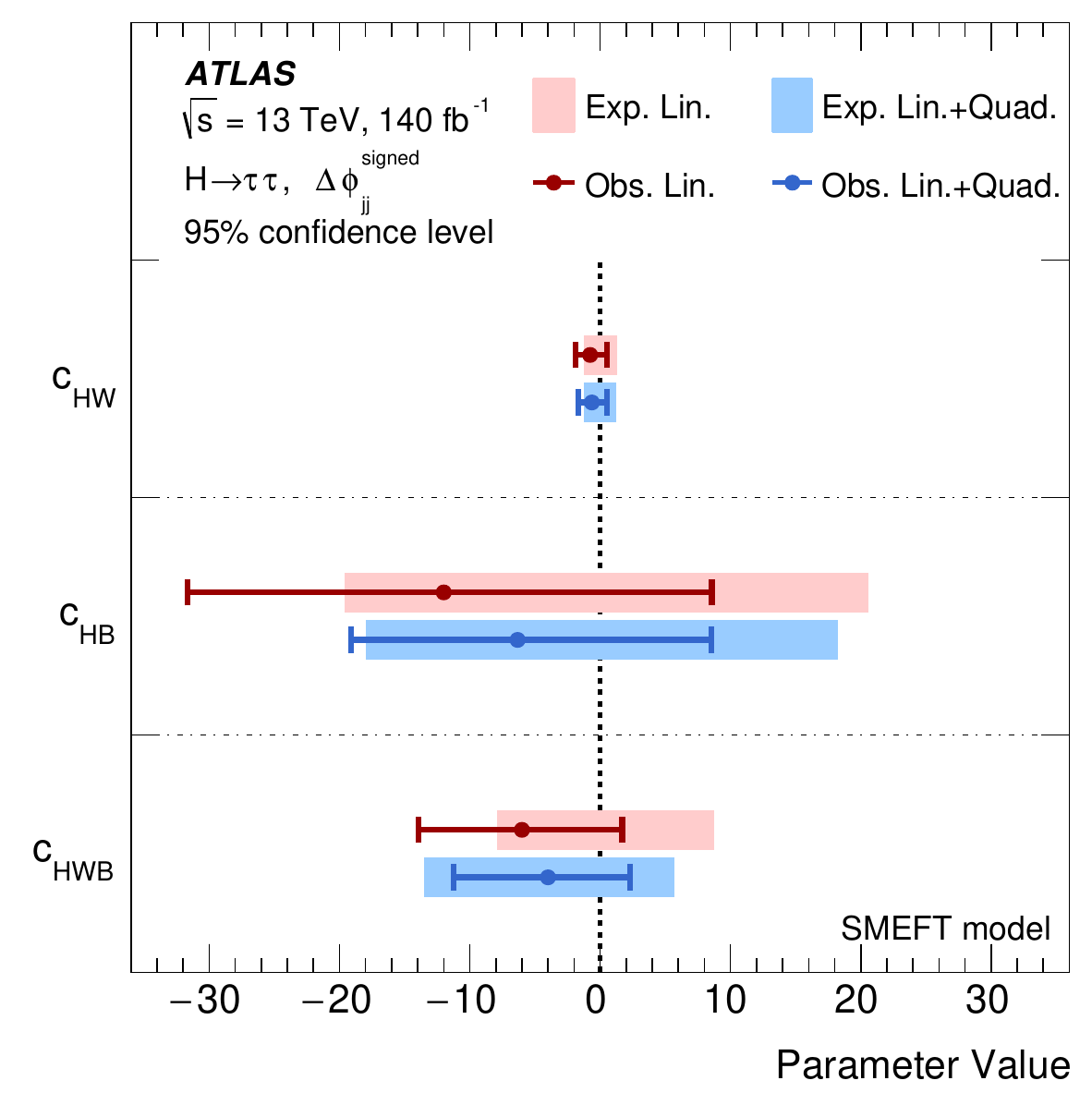} &
\includegraphics[width=0.4\linewidth]{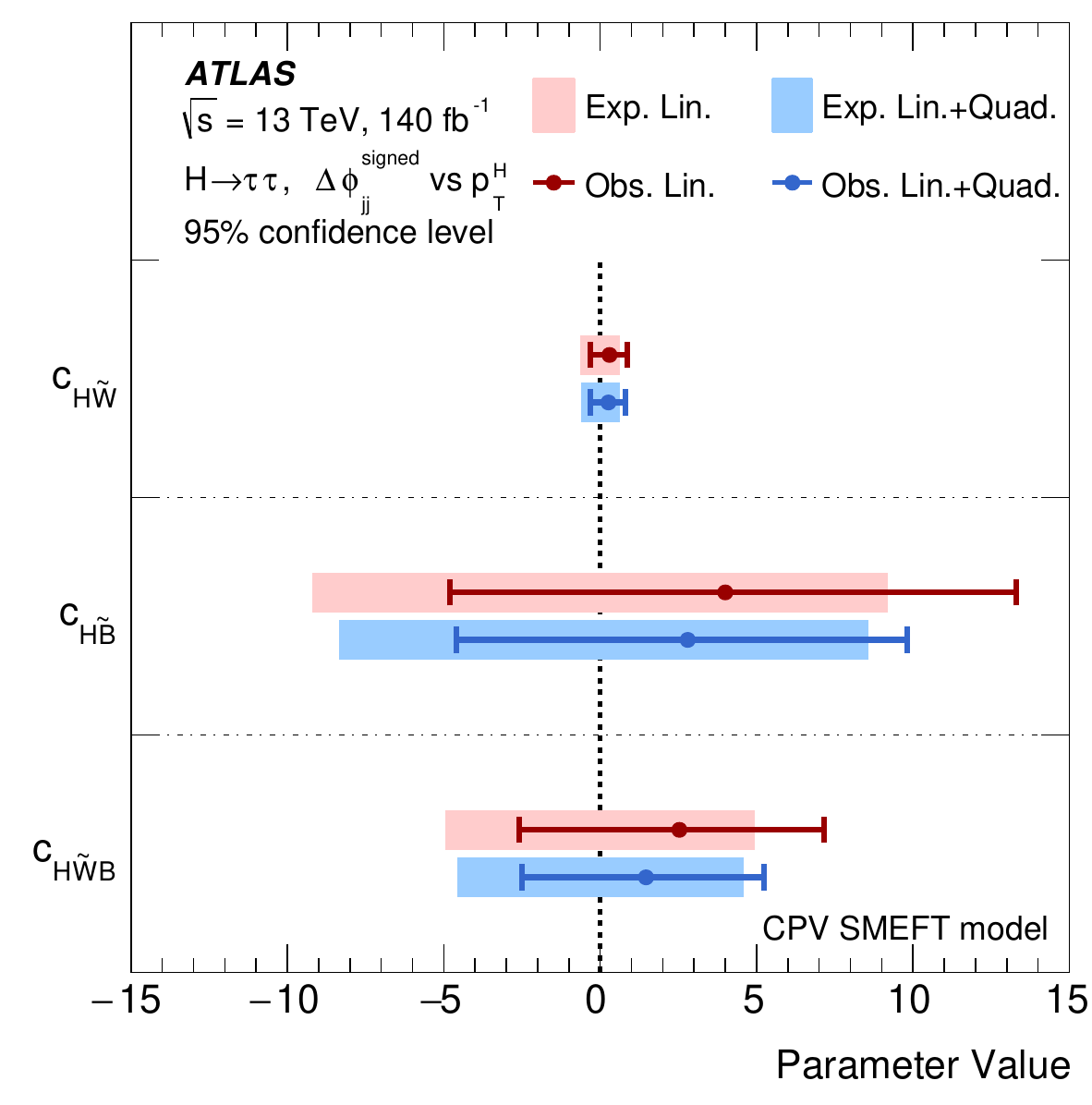} \\
(a) &
(b) \\
\end{tabular}
\caption{
The 95\% expected and observed confidence intervals for each of the six considered Wilson coefficients. Each coefficient is treated individually and both the linear and linear~$+$~quadratic models are considered when evaluating the sensitivity to the terms that enter at $\mathcal{O}(\Lambda^{-4})$. For the (a) CP-even operators the \dPhiJJsign distribution is used to extract the confidence interval, while for (b) the CP-odd operators the \dPhiJJsign vs \pTH distribution is used. The limits are computed at a new-physics scale $\Lambda=1$~TeV.
\label{fig:diffEFT:1Dresults}}
\end{center}
\end{figure}

Firstly considered is the scenario where only one of the Wilson coefficients is non-zero and the rest are neglected. The unfolded data are used along with the theoretical dependence of the cross-section on the Wilson coefficients to extract the best-fit value of each of the six considered Wilson coefficients. The same profile-likelihood formalism as presented in \cref{sec:differential_results:fitModel} is used with  only one degree of freedom introduced, corresponding to the floating Wilson coefficient, and all other Wilson coefficients are set to zero. A scan of the negative log-likelihood described in \cref{sec:differential_results:fitModel} is performed  for different values of each Wilson coefficient, with the cross-section in each bin fixed to the value predicted by the SMEFT. By considering the change in the negative log-likelihood from its minimum value, the 95\% confidence interval for each Wilson coefficient is found. This is done for only the linear term as well as for both the linear and quadratic terms. For the Wilson coefficients associated with CP-even operators the \dPhiJJsign distribution is used, while for those associated with CP-odd operators the \dPhiJJsign vs \pTH distribution is used. These are the distributions that give the best expected precision. \cref{fig:diffEFT:1Dresults} shows the expected and observed confidence intervals for each of the six coefficients individually. Other than for the $c_{HWB}$ coefficient, the intervals considering only the linear term are very similar to the one when both the linear and quadratic terms are considered. Therefore, the approach of using an EFT only considering dimension-six operators is valid for most of the operators considered in the analysis. The observed (expected) intervals for $c_{HW}$ and $c_{H\tilde{W}}$ are $[-1.90, +0.51]$ ($[-1.19, +1.21]$) and $[-0.31, +0.88]$ ($[-0.60, +0.60]$) in the linear scenario and are the most stringent among the considered coefficients. The constraints on $c_{H\tilde{W}}$ are the tightest to date.

\begin{figure}[htb!]
\begin{center}
\begin{tabular}{cc}
\includegraphics[width=0.45\linewidth]{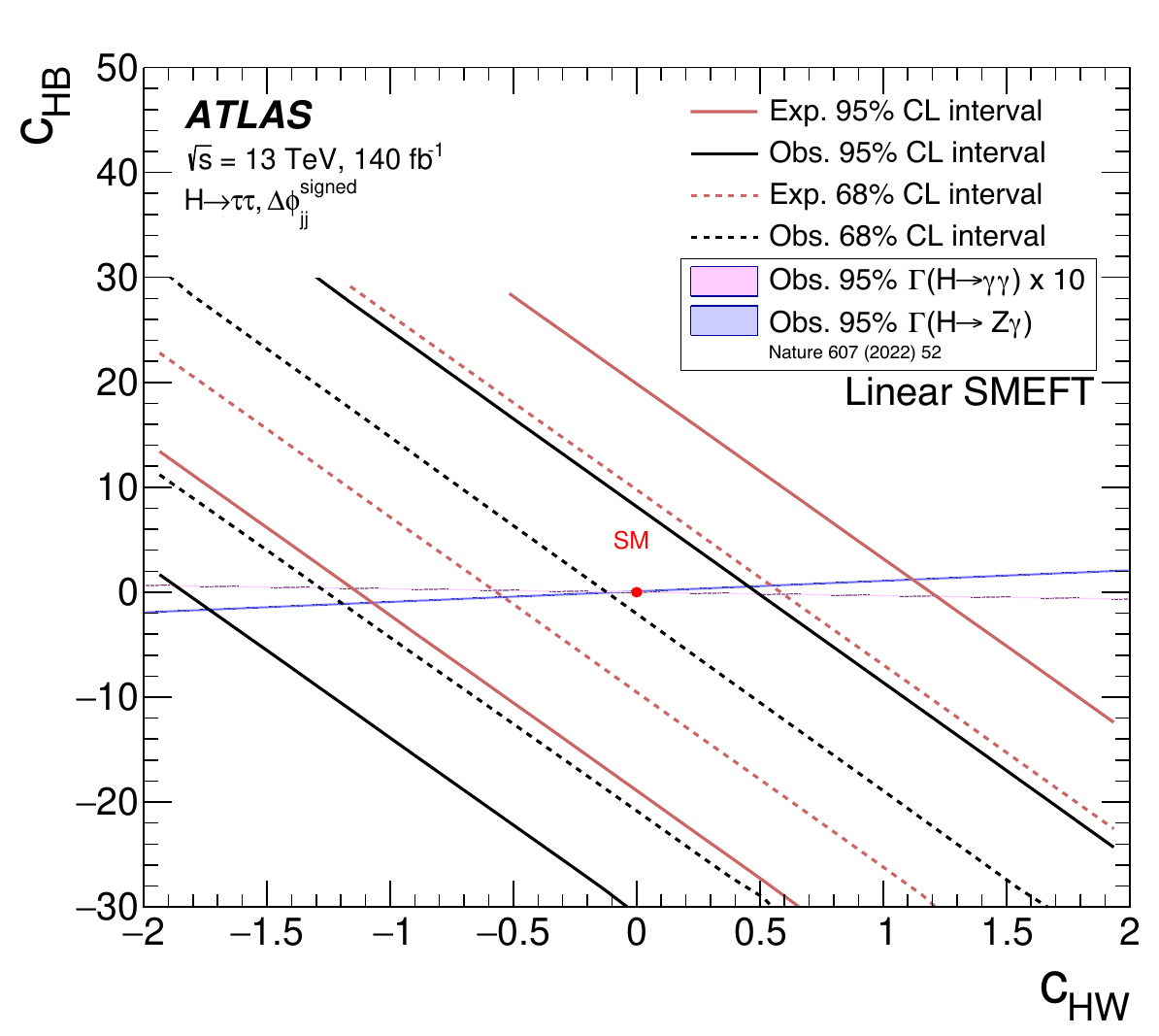}  &
\includegraphics[width=0.45\linewidth]{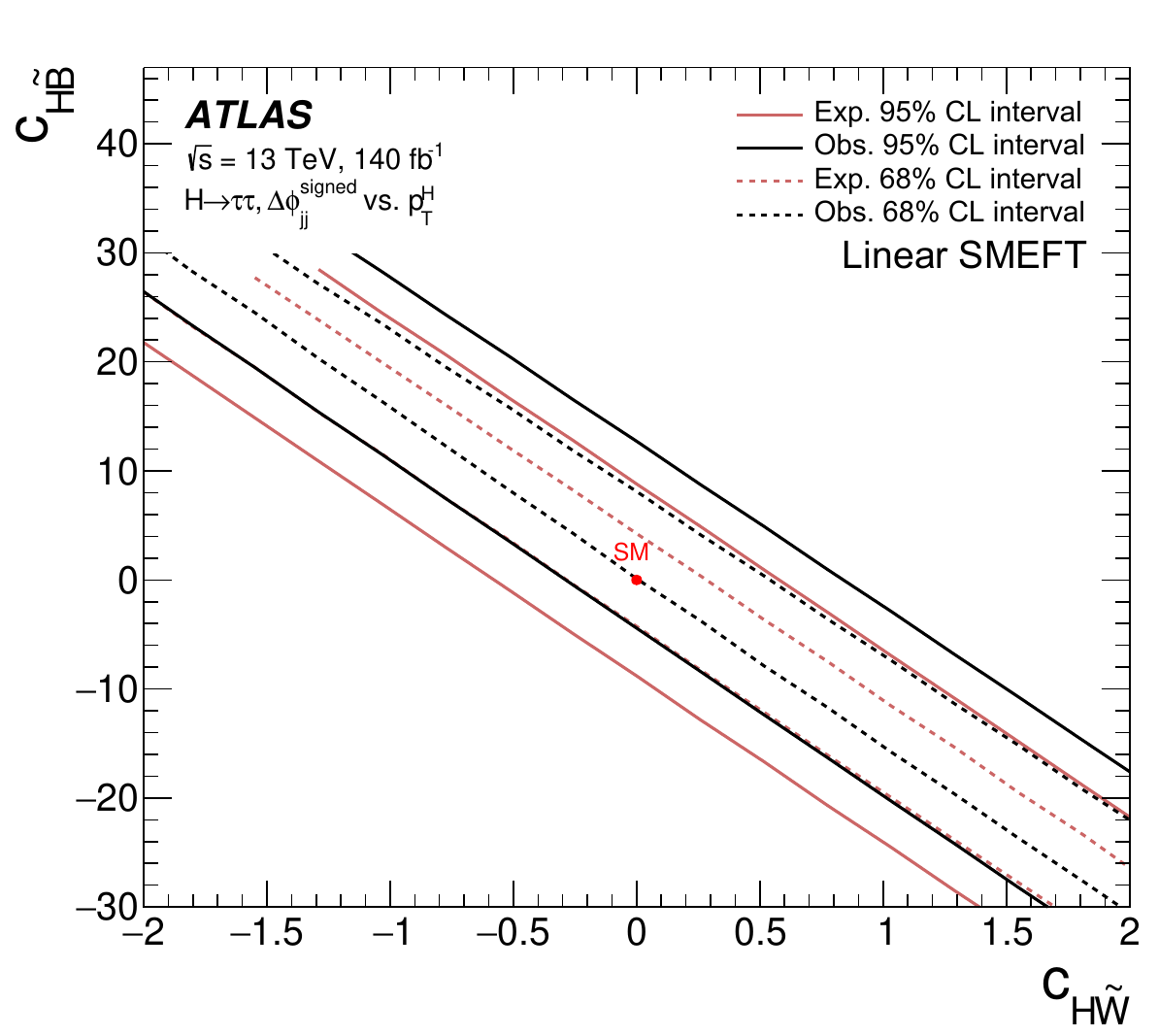} \\
(a) &
(b) \\
\multicolumn{2}{c}{\includegraphics[width=0.45\linewidth]{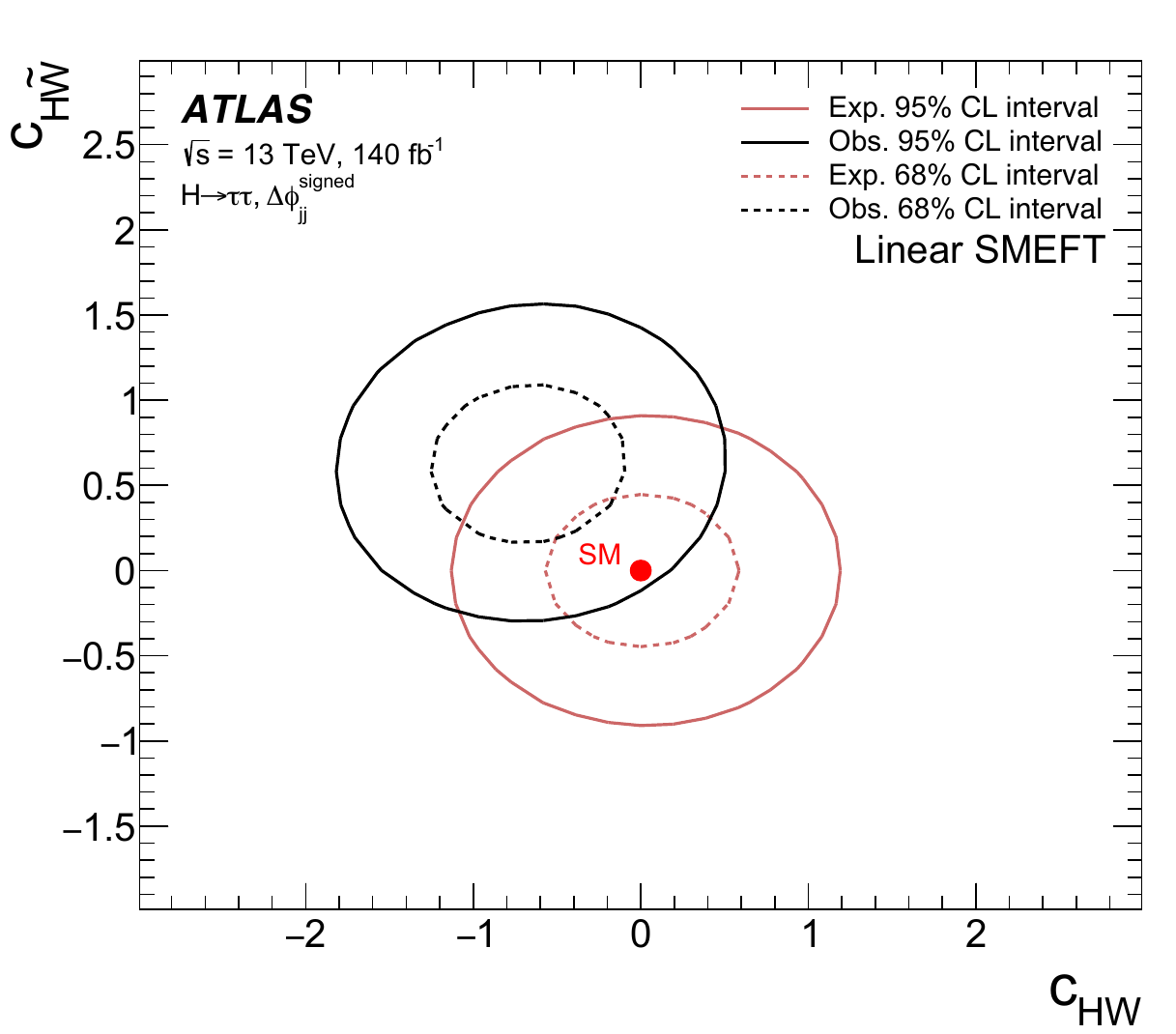}} \\
\multicolumn{2}{c}{(c)} \\
\end{tabular}
\caption{
Two-dimensional observed and expected limits at 68\% and 95\% confidence level obtained from various combinations of two Wilson coefficients using only the SM--dimension-6 interference in the SMEFT basis: (a) $c_{HW}$ vs $c_{HB}$, (b) $c_{H\tilde{W}}$ vs $c_{H\tilde{B}}$ and (c) $c_{HW}$ vs $c_{H\tilde{W}}$. The limits are computed at a new-physics scale $\Lambda=1$~TeV. In (a) and (c), the \dPhiJJsign distribution is used, while in the purely CP-odd case (b), the \dPhiJJsign vs \pTH distribution is used. To illustrate the complementarity of this measurement with other Higgs boson studies, in (a) the limits from measurements in the $H\rightarrow\gamma\gamma$ and $H\rightarrow Z\gamma$~\cite{HIGG-2021-23} interpreted in the SMEFT \cite{Dawson:2018liq,Dawson:2018pyl} are also shown. The interval resulting from the $H\rightarrow\gamma\gamma$ measurement has been increased by a factor of ten for visibility.
\label{fig:diff:2Dresults}}
\end{center}
\end{figure}

The operators considered are closely related and new physics could result in contributions from multiple operators~\cite{deBlas:2017xtg} such that it is also interesting to explore the case where two of the operators are non-zero. This proceeds in the same way by scanning the change in negative log-likelihood across a plane of two of the Wilson coefficients while the others are set to zero. Only terms which enter at $\mathcal{O}(\Lambda^{-2})$ are considered. The first plane considered, in \cref{fig:diff:2Dresults}(a), is $c_{HW}$ vs $c_{HB}$ where the \dPhiJJsign distribution is used to perform the measurement. The effects of these two coefficients are very similar such that there is a `flat direction' where the effects of the two Wilson coefficients cancel each other out and there is no sensitivity from the analysis. As one moves away from this direction the analysis gains sensitivity and can constrain the allowed values of the coefficients. %
In Ref.~\cite{HIGG-2021-23} a fit across multiple Higgs analyses measured the Higgs boson's decay width into two photons or into a $Z$ boson and a photon. These two decay channels only proceed through loop diagrams in the SM so they are very sensitive to BSM physics contributions which have been calculated at NLO~\cite{Dawson:2018liq,Dawson:2018pyl}.
These channels also have `flat directions' where the $\mathcal{O}(\Lambda^{-2})$ contributions from the two operators cancel each other out. However, these are different from the `flat direction' in this analysis, showing the complementarity of the different types of measurement and that multiple measurements are required to constrain the multi-dimensional EFT parameter space. It should also be noted that while in the displayed plane $c_{HWB}$ is set to zero, both this analysis and the two decay widths also have a dependence on this parameter such that in three dimensions any two measurements will still have a direction in which neither is sensitive.
The second considered plane, in \cref{fig:diff:2Dresults}(b), is $c_{H\tilde{W}}$ vs $c_{H\tilde{B}}$, where the \dPhiJJsign vs \pTH distribution is used. This also shows a `flat direction' and also tighter constraints than the first plane because of the two-dimensional distribution used. There are no constraints from Higgs boson decay widths in this case, due to the CP-odd nature of the operators. Finally, the plane of $c_{HW}$ vs $c_{H\tilde{W}}$  in \cref{fig:diff:2Dresults}(c) is considered using the \dPhiJJsign distribution. In this case the two operators introduce distinct shape differences to the distribution such that there are no `flat directions'. In all cases the observed 95\% confidence limits agree well with the expected confidence limits such that there is no evidence of BSM physics.

%
%
%
%

%
%
%
%

%

%
%
%

\FloatBarrier
\section{Conclusion}
\label{sec:conclusion}

Measurements of Higgs boson production cross-sections in the $\tau$-lepton-pair decay channel are presented.  The analysis uses 140\,fb$^{-1}$ of proton--proton collisions at $\sqrt{s}=13$~TeV recorded by the ATLAS detector at the LHC and supersedes the previous analysis of the same dataset. Two different types of measurement were performed: cross-section measurements of the four main production processes in the STXS framework, including eight kinematic bins for VBF production, and unfolded differential measurements in a region of phase space enhanced in VBF production. The latter are also used to derive 95\% confidence intervals for dimension-six operators in the SMEFT formalism. %
Many of these measurements are limited by the size of the analyzed data sample.

Cross-section measurements for the four main Higgs boson production modes (gluon--gluon fusion, vector-boson fusion, associated production with a vector boson, and associated production with a top--antitop quark pair) are found to agree well with the SM predictions.
In particular, the ratio of the measured cross-section to the SM prediction for VBF Higgs boson production is found to be $1.04^{+0.19}_{-0.17}$, representing the most precise measurement of this production mode across all Higgs boson results. %

Eighteen bins of the STXS framework were measured, with an overall $p$-value of 6\% for compatibility with the SM prediction, demonstrating reasonable agreement.
For VBF production, higher precision is achieved in the higher $\pTH$~and higher $\mjj$~regions because of their higher signal purity.
The ratio of the measured cross-section to the SM prediction for VBF production with $\pTH>200~\GeV$ ($\pTH<200~\GeV$) and $\mjj>1.5~\TeV$ is ${1.29}^{+0.39}_{-0.34}$ (${0.12}^{+0.34}_{-0.33}$). This is the first measurement for the higher \pTH~range, and the most precise for the lower \pTH~range, in this corner of phase space.

Unfolded differential cross-sections for Higgs boson production in a region of phase space dominated by VBF production are also presented.  Four distributions are unfolded to particle level: \pTH, \pTj, \dPhiJJsign, and \dPhiJJsign vs \pTH.
All of these distributions show good agreement with the SM predictions, with typical uncertainties of 30\%--50\%, dominated by the statistical uncertainty of the data.

These distributions are sensitive to BSM physics and a SMEFT interpretation is presented. Six operators were considered, all affecting the VBF Higgs boson production vertex, and the associated Wilson coefficients are $c_{HW}, c_{HB}, c_{HWB}$, and their CP-odd counterparts. For each coefficient, 95\% confidence intervals were derived for the interference term alone and for both the interference and purely BSM contributions. The constraints on the CP-odd Wilson coefficient $c_{H\tilde{W}}$ in the former case, $[-0.31, +0.88]$, are the strongest to date from any channel.
Two-dimensional 95\% confidence intervals were also derived, demonstrating unique sensitivity for these operators compared to other Higgs boson analyses.

%


%

%

%
\section*{Acknowledgements}
%

%
%

%
%

We thank CERN for the very successful operation of the LHC and its injectors, as well as the support staff at
CERN and at our institutions worldwide without whom ATLAS could not be operated efficiently.

The crucial computing support from all WLCG partners is acknowledged gratefully, in particular from CERN, the ATLAS Tier-1 facilities at TRIUMF/SFU (Canada), NDGF (Denmark, Norway, Sweden), CC-IN2P3 (France), KIT/GridKA (Germany), INFN-CNAF (Italy), NL-T1 (Netherlands), PIC (Spain), RAL (UK) and BNL (USA), the Tier-2 facilities worldwide and large non-WLCG resource providers. Major contributors of computing resources are listed in Ref.~\cite{ATL-SOFT-PUB-2023-001}.

We gratefully acknowledge the support of ANPCyT, Argentina; YerPhI, Armenia; ARC, Australia; BMWFW and FWF, Austria; ANAS, Azerbaijan; CNPq and FAPESP, Brazil; NSERC, NRC and CFI, Canada; CERN; ANID, Chile; CAS, MOST and NSFC, China; Minciencias, Colombia; MEYS CR, Czech Republic; DNRF and DNSRC, Denmark; IN2P3-CNRS and CEA-DRF/IRFU, France; SRNSFG, Georgia; BMBF, HGF and MPG, Germany; GSRI, Greece; RGC and Hong Kong SAR, China; ISF and Benoziyo Center, Israel; INFN, Italy; MEXT and JSPS, Japan; CNRST, Morocco; NWO, Netherlands; RCN, Norway; MNiSW, Poland; FCT, Portugal; MNE/IFA, Romania; MSTDI, Serbia; MSSR, Slovakia; ARIS and MVZI, Slovenia; DSI/NRF, South Africa; MICIU/AEI, Spain; SRC and Wallenberg Foundation, Sweden; SERI, SNSF and Cantons of Bern and Geneva, Switzerland; NSTC, Taipei; TENMAK, T\"urkiye; STFC/UKRI, United Kingdom; DOE and NSF, United States of America.

Individual groups and members have received support from BCKDF, CANARIE, CRC and DRAC, Canada; CERN-CZ, FORTE and PRIMUS, Czech Republic; COST, ERC, ERDF, Horizon 2020, ICSC-NextGenerationEU and Marie Sk{\l}odowska-Curie Actions, European Union; Investissements d'Avenir Labex, Investissements d'Avenir Idex and ANR, France; DFG and AvH Foundation, Germany; Herakleitos, Thales and Aristeia programmes co-financed by EU-ESF and the Greek NSRF, Greece; BSF-NSF and MINERVA, Israel; NCN and NAWA, Poland; La Caixa Banking Foundation, CERCA Programme Generalitat de Catalunya and PROMETEO and GenT Programmes Generalitat Valenciana, Spain; G\"{o}ran Gustafssons Stiftelse, Sweden; The Royal Society and Leverhulme Trust, United Kingdom.

In addition, individual members wish to acknowledge support from Armenia: Yerevan Physics Institute (FAPERJ); CERN: European Organization for Nuclear Research (CERN PJAS); Chile: Agencia Nacional de Investigaci\'on y Desarrollo (FONDECYT 1230812, FONDECYT 1230987, FONDECYT 1240864); China: Chinese Ministry of Science and Technology (MOST-2023YFA1605700, MOST-2023YFA1609300), National Natural Science Foundation of China (NSFC - 12175119, NSFC 12275265, NSFC-12075060); Czech Republic: Czech Science Foundation (GACR - 24-11373S), Ministry of Education Youth and Sports (FORTE CZ.02.01.01/00/22\_008/0004632), PRIMUS Research Programme (PRIMUS/21/SCI/017); EU: H2020 European Research Council (ERC - 101002463); European Union: European Research Council (ERC - 948254, ERC 101089007), Horizon 2020 Framework Programme (MUCCA - CHIST-ERA-19-XAI-00), European Union, Future Artificial Intelligence Research (FAIR-NextGenerationEU PE00000013), Italian Center for High Performance Computing, Big Data and Quantum Computing (ICSC, NextGenerationEU); France: Agence Nationale de la Recherche (ANR-20-CE31-0013, ANR-21-CE31-0013, ANR-21-CE31-0022, ANR-22-EDIR-0002), Investissements d'Avenir Labex (ANR-11-LABX-0012); Germany: Baden-Württemberg Stiftung (BW Stiftung-Postdoc Eliteprogramme), Deutsche Forschungsgemeinschaft (DFG - 469666862, DFG - CR 312/5-2); Italy: Istituto Nazionale di Fisica Nucleare (ICSC, NextGenerationEU), Ministero dell'Università e della Ricerca (PRIN - 20223N7F8K - PNRR M4.C2.1.1); Japan: Japan Society for the Promotion of Science (JSPS KAKENHI JP22H01227, JSPS KAKENHI JP22H04944, JSPS KAKENHI JP22KK0227, JSPS KAKENHI JP23KK0245); Netherlands: Netherlands Organisation for Scientific Research (NWO Veni 2020 - VI.Veni.202.179); Norway: Research Council of Norway (RCN-314472); Poland: Ministry of Science and Higher Education (IDUB AGH, POB8, D4 no 9722), Polish National Agency for Academic Exchange (PPN/PPO/2020/1/00002/U/00001), Polish National Science Centre (NCN 2021/42/E/ST2/00350, NCN OPUS nr 2022/47/B/ST2/03059, NCN UMO-2019/34/E/ST2/00393, NCN \& H2020 MSCA 945339, UMO-2020/37/B/ST2/01043, UMO-2021/40/C/ST2/00187, UMO-2022/47/O/ST2/00148, UMO-2023/49/B/ST2/04085, UMO-2023/51/B/ST2/00920); Slovenia: Slovenian Research Agency (ARIS grant J1-3010); Spain: Generalitat Valenciana (Artemisa, FEDER, IDIFEDER/2018/048), Ministry of Science and Innovation (MCIN \& NextGenEU PCI2022-135018-2, MICIN \& FEDER PID2021-125273NB, RYC2019-028510-I, RYC2020-030254-I, RYC2021-031273-I, RYC2022-038164-I), PROMETEO and GenT Programmes Generalitat Valenciana (CIDEGENT/2019/027); Sweden: Carl Trygger Foundation (Carl Trygger Foundation CTS 22:2312), Swedish Research Council (Swedish Research Council 2023-04654, VR 2018-00482, VR 2022-03845, VR 2022-04683, VR 2023-03403, VR grant 2021-03651), Knut and Alice Wallenberg Foundation (KAW 2018.0157, KAW 2018.0458, KAW 2019.0447, KAW 2022.0358); Switzerland: Swiss National Science Foundation (SNSF - PCEFP2\_194658); United Kingdom: Leverhulme Trust (Leverhulme Trust RPG-2020-004), Royal Society (NIF-R1-231091); United States of America: U.S. Department of Energy (ECA DE-AC02-76SF00515), Neubauer Family Foundation.

%
%


%
\clearpage
\printbibliography
\clearpage
 
\begin{flushleft}
\hypersetup{urlcolor=black}
{\Large The ATLAS Collaboration}

\bigskip

\AtlasOrcid[0000-0002-6665-4934]{G.~Aad}$^\textrm{\scriptsize 104}$,
\AtlasOrcid[0000-0001-7616-1554]{E.~Aakvaag}$^\textrm{\scriptsize 17}$,
\AtlasOrcid[0000-0002-5888-2734]{B.~Abbott}$^\textrm{\scriptsize 123}$,
\AtlasOrcid[0000-0002-0287-5869]{S.~Abdelhameed}$^\textrm{\scriptsize 119a}$,
\AtlasOrcid[0000-0002-1002-1652]{K.~Abeling}$^\textrm{\scriptsize 56}$,
\AtlasOrcid[0000-0001-5763-2760]{N.J.~Abicht}$^\textrm{\scriptsize 50}$,
\AtlasOrcid[0000-0002-8496-9294]{S.H.~Abidi}$^\textrm{\scriptsize 30}$,
\AtlasOrcid[0009-0003-6578-220X]{M.~Aboelela}$^\textrm{\scriptsize 45}$,
\AtlasOrcid[0000-0002-9987-2292]{A.~Aboulhorma}$^\textrm{\scriptsize 36e}$,
\AtlasOrcid[0000-0001-5329-6640]{H.~Abramowicz}$^\textrm{\scriptsize 155}$
% ...
\bigskip
\\

% ...
$^{17}$Department for Physics and Technology, University of Bergen, Bergen; Norway.\\
% ...
$^{30}$Physics Department, Brookhaven National Laboratory, Upton NY; United States of America.\\
% ...
$^{36}$$^{(a)}$Facult\'e des Sciences Ain Chock, Universit\'e Hassan II de Casablanca;$^{(b)}$Facult\'{e} des Sciences, Universit\'{e} Ibn-Tofail, K\'{e}nitra;$^{(c)}$Facult\'e des Sciences Semlalia, Universit\'e Cadi Ayyad, LPHEA-Marrakech;$^{(d)}$LPMR, Facult\'e des Sciences, Universit\'e Mohamed Premier, Oujda;$^{(e)}$Facult\'e des sciences, Universit\'e Mohammed V, Rabat;$^{(f)}$Institute of Applied Physics, Mohammed VI Polytechnic University, Ben Guerir; Morocco.\\
% ...
$^{45}$Physics Department, Southern Methodist University, Dallas TX; United States of America.\\
% ...
$^{50}$Fakult\"{a}t Physik , Technische Universit{\"a}t Dortmund, Dortmund; Germany.\\
% ...
$^{56}$II. Physikalisches Institut, Georg-August-Universit\"{a}t G\"ottingen, G\"ottingen; Germany.\\
% ...
$^{104}$CPPM, Aix-Marseille Universit\'e, CNRS/IN2P3, Marseille; France.\\
% ...
$^{119}$$^{(a)}$New York University Abu Dhabi, Abu Dhabi;$^{(b)}$United Arab Emirates University, Al Ain; United Arab Emirates.\\
% ...
$^{123}$Homer L. Dodge Department of Physics and Astronomy, University of Oklahoma, Norman OK; United States of America.\\
% ...
$^{155}$Raymond and Beverly Sackler School of Physics and Astronomy, Tel Aviv University, Tel Aviv; Israel.\\
% ...
$^{a}$ Also Affiliated with an institute covered by a cooperation agreement with CERN.\\
$^{b}$ Also at An-Najah National University, Nablus; Palestine.\\
$^{c}$ Also at Borough of Manhattan Community College, City University of New York, New York NY; United States of America.\\
$^{d}$ Also at Center for High Energy Physics, Peking University; China.\\
$^{e}$ Also at Center for Interdisciplinary Research and Innovation (CIRI-AUTH), Thessaloniki; Greece.\\
$^{f}$ Also at CERN, Geneva; Switzerland.\\
% ...

\end{flushleft}


%

%
%

%

%
%
%
%

\end{document}